\documentclass[conference]{IEEEtran}
\IEEEoverridecommandlockouts

\usepackage{tikz}
\usepackage{mathptmx} % This is Times font

\usepackage{amsmath}
\usepackage[normalem]{ulem}
\usepackage{amssymb}
\usepackage{multicol}
\usepackage{graphicx}
\usepackage{pifont}
\usepackage[noend]{algorithmic}
\usepackage[ruled, linesnumbered]{algorithm2e}
\usepackage{enumitem}
\usepackage{multirow}
\usepackage{xcolor}
\usepackage{color}
\usepackage{listings}
\usepackage{float}
\usepackage{ragged2e}
\usepackage{xspace}
\usepackage{soul}
\usepackage{lipsum}
\usepackage{makecell}
\usepackage{mathtools}
\usepackage{subcaption}
\usepackage{fancyhdr}
\usepackage[normalem]{ulem}
\usepackage[hyphens]{url}
\usepackage[sort,nocompress]{cite}
\usepackage[final]{microtype}
\usepackage[keeplastbox]{flushend}
\usepackage{xcolor}
\usepackage{listings}
\usepackage{multirow}
\usepackage{wrapfig}
\usepackage{etoolbox}

\usepackage{fancyhdr}

\usepackage[bookmarks=true,breaklinks=true,letterpaper=true,colorlinks,linkcolor=black,citecolor=blue,urlcolor=black]{hyperref}
\usepackage[colorinlistoftodos,prependcaption,textsize=small]{todonotes}
\newcommand{\hlc}[2][white]{{%
    \colorlet{foo}{#1}%
    \sethlcolor{foo}\hl{#2}}%
}

\newcommand{\proj}{\benchmark{TW}\xspace}
\newcommand{\Fig}[1]{Fig.~\ref{#1}}

\newcommand{\Sec}[1]{Sec.~\ref{#1}}

\newcommand{\benchmark}[1]{{\texttt{#1}}}
\renewcommand{\paragraph}[1]{\vspace*{0.15cm}\noindent\textbf{#1}\hspace*{.1cm}}

\newcommand*\circled[1]{\tikz[baseline=(char.base)]{
               \node[shape=circle,fill,inner sep=0.6pt] (char) {\textcolor{white}{#1}};}}

\def\BibTeX{{\rm B\kern-.05em{\sc i\kern-.025em b}\kern-.08em
    T\kern-.1667em\lower.7ex\hbox{E}\kern-.125emX}}
\title{Accelerating Sparse DNN Models without Hardware-Support via Tile-Wise Sparsity}

\begin{document}
%-------------------------------------------------------------------------------
\makeatletter
\newcommand{\linebreakand}{%
  \end{@IEEEauthorhalign}
  \hfill\mbox{}\par
  \mbox{}\hfill\begin{@IEEEauthorhalign}
}
\makeatother

\author{\IEEEauthorblockN{Cong Guo$^*$\thanks{$^*$Contribution during his internship at NVIDIA.}}
\IEEEauthorblockA{
\textit{Shanghai Jiao Tong University}\\
guocong@sjtu.edu.cn}
\and

\IEEEauthorblockN{Bo Yang Hsueh}
\IEEEauthorblockA{
\textit{NVIDIA}\\
bhsueh@nvidia.com}
\and

\IEEEauthorblockN{Jingwen Leng$^\S$}
\IEEEauthorblockA{
\textit{Shanghai Jiao Tong University}\\ \textit{Shanghai Qi Zhi Institute}\\
leng-jw@sjtu.edu.cn}

\linebreakand

\IEEEauthorblockN{Yuxian Qiu}
\IEEEauthorblockA{
\textit{Shanghai Jiao Tong University}\\
qiuyuxian@sjtu.edu.cn}
\and

\IEEEauthorblockN{Yue Guan}
\IEEEauthorblockA{
\textit{Shanghai Jiao Tong University}\\
bonboru@sjtu.edu.cn}
\and

\IEEEauthorblockN{Zehuan Wang}
\IEEEauthorblockA{
  \textit{NVIDIA}\\
zehuanw@nvidia.com}
\and

\IEEEauthorblockN{Xiaoying Jia}
\IEEEauthorblockA{
  \textit{NVIDIA}\\
irishcoffee1006@gmail.com}
\and

\linebreakand

\IEEEauthorblockN{Xipeng Li}
\IEEEauthorblockA{
  \textit{NVIDIA}\\
xipengl@nvidia.com}
\and

\IEEEauthorblockN{Minyi Guo$^{\S}$\thanks{${^\S}$Jingwen Leng and Minyi Guo are corresponding authors of this paper.}}
\IEEEauthorblockA{
\textit{Shanghai Jiao Tong University}\\
\textit{Shanghai Qi Zhi Institute}\\
guo-my@cs.sjtu.edu.cn}
\and

\IEEEauthorblockN{Yuhao Zhu}
\IEEEauthorblockA{
\textit{University of Rochester}\\
yzhu@rochester.edu}
}

\maketitle

\thispagestyle{fancy}
\lhead{}
\rhead{}
\chead{}
\lfoot{\footnotesize{
SC20, November 9-19, 2020, Is Everywhere We Are
\newline 978-1-7281-9998-6/20/\$31.00 \copyright 2020 IEEE}}
\rfoot{}
\cfoot{}
\renewcommand{\headrulewidth}{0pt}
\renewcommand{\footrulewidth}{0pt}

\begin{abstract} 

Network pruning can reduce the high computation cost of deep neural network (DNN) models. However, to maintain their accuracies, sparse models often carry randomly-distributed weights, leading to irregular computations.
Consequently, sparse models cannot achieve meaningful speedup on commodity hardware (e.g., GPU) built for dense matrix computations.
As such, prior works usually modify or design completely new sparsity-optimized architectures for exploiting sparsity.
We propose an algorithm-software co-designed pruning method that achieves latency speedups on existing dense architectures.
Our work builds upon the insight that the matrix multiplication generally breaks the large matrix into multiple smaller tiles for parallel execution. 
We propose a tiling-friendly ``tile-wise'' sparsity pattern, which maintains a regular pattern at the tile level for efficient execution but allows for irregular, arbitrary pruning at the global scale to maintain the high accuracy.
We implement and evaluate the sparsity pattern on GPU tensor core, achieving a $1.95\times$ speedup over the dense model.

\end{abstract}

% %\begin{IEEEkeywords}
%     BERT, Pruning, GPU, \proj
% %\end{IEEEkeywords}
\section{Introduction}\label{sec:introduction}

\hlc{Deep} neural network (DNN) models have achieved and even surpassed human-level accuracy in important domains~\cite{dnnvshuman}. 
For instance, transformer-based models~\cite{46201} in natural language processing (NLP) such as BERT~\cite{devlin2018bert} have dominated the accuracy in various NLP tasks and have been used in the Google's new search algorithm~\cite{googlebert}. Despite their high accuracies, DNN models also have significant computational cost, both in training and inference.
The large NLP model GPT-2~\cite{radford2019language} has 1.5~billion parameters and takes roughly a week to train on 32 TPUv3 chips and costs over \$40,000 on Google's cloud TPU platform~\cite{bertcost2}. 
The inference latency of modern DNN models could also be excessively high due to the enormous computation cost and memory usage~\cite{patdnn}.

One particularly effective and promising approach to reduce the DNN latency is pruning~\cite{lecun1990optimal, han2015learning}, which exploits the inherent redundancy in the DNN models to transform the original, dense model to a sparse model by iteratively removing ``unimportant'' weight elements and retraining the model to recover its accuracy loss. In the end, the sparse model has fewer parameters and, theoretically, less computation cost.

The primary challenge in network pruning is how to balance the model accuracy and execution efficiency.
Such a balance is fundamentally affected by the 
\textit{sparsity pattern} that a pruning approach enforces. Intuitively, a stronger constraint on the sparsity pattern forces certain weights to be pruned and, thus, leads to lower accuracy, and vice-versa.
The most fine-grained pruning approach leads to the so-called element-wise (\benchmark{EW}) sparsity pattern, which prunes weight elements individually and independently, solely by their importance scores~\cite{han2015learning}.
In other words, \benchmark{EW} imposes \textit{no} constraints on the sparsity pattern and can remove any weight element, leading to  the minimal model accuracy degradation.
However, the pruned sparse model also introduces irregular memory accesses that are unfriendly on commodity architectures, e.g., GPUs~\cite{leng13gpuwattch,rhu2013lamar}.
As a result, \benchmark{EW}-based sparse DNN models usually run slower than the unpruned dense models on these architectures~\cite{hill2017deftnn}.

To realize the acceleration potential of sparse DNN models, researchers have proposed to co-design the sparsity pattern with hardware support.
For instance, many architects have proposed various specialized accelerator designs~\cite{eie, scnn} to exploit the zeros in the aforementioned \benchmark{EW} pattern for latency reduction.
Similarly, prior work proposes the vector-wise pattern~\cite{yao2019balanced} that divides a weight column to groups and prunes the same number of elements in a group. 
This sparsity pattern requires the new hardware or the modification of existing hardware~\cite{zhu2019sparse, columncombining}.
In summary, these approaches lead to sparse memory accesses and computation patterns that require hardware support to be effective, and thus cannot leverage commodity DNN accelerators such TPU~\cite{jouppi2017datacenter} and Volta tensor core~\cite{volta2017whitepaper, tensorcoreispass}.

In this work, we propose a novel algorithm that can accelerate sparse DNN models on commodity DNN accelerators without hardware modification. Our key observation is that virtually all of today's DNN accelerators implement dense general matrix multiplication (GEMM)~\cite{chellapilla2006high} operations. GEMM-based accelerators~\cite{volta2017whitepaper, jouppi2017datacenter, guo2020balancing, qin2020sigma} are dominant owing to their wide applicability: convolution operations that dominate computer vision models are lowered to the GEMM operation, and NLP models are naturally equivalent to the GEMM operation. Examples include NVIDIA's tensor core~\cite{volta2017whitepaper} and Google's TPU~\cite{jouppi2017datacenter} mentioned above. We propose a new pruning algorithm, which enforces a particular sparsity pattern on pruned models to directly leverage existing GEMM accelerators without modifying the microarchitectures.

In particular, our work exploits the key insight that the matrix multiplication on existing dense GEMM accelerators adopts the tiling approach, which breaks the large matrix into multiple smaller tiles for parallel execution. We propose a tiling-friendly sparsity pattern called \emph{tile sparsity} (or \benchmark{TW}), which maintains a regular sparsity pattern at the tile level for efficient execution but allows for irregular, arbitrary pruning at a global scale through non-uniform tile sizes to maintain high model accuracies.

To exploit the \texttt{TW} sparsity, we first divide the entire matrix into multiple tiles as in conventional tiled GEMM. We then prune the \textit{entire rows or columns} of each tile according to the collective importance scores of each row and column.
In our sparsity pattern, the tile size dictates the trade-off between model accuracy and execution efficiency.
At one extreme where the tile size equals one, our \texttt{TW} sparsity is equivalent to the \texttt{EW} sparsity.
At the other extreme where the tile size is the same as the matrix size, \texttt{TW} pruning is equivalent to the global structural pruning that prunes the entire row or column~\cite{hill2017deftnn}.

Building on top of \texttt{TW}, we further propose a \emph{hybrid} sparsity pattern that overlays the most fine-grained \benchmark{EW} sparsity pattern on top of the \texttt{TW} sparsity.
With a small fraction of \benchmark{EW} ({e.g., 1.5\%}), the hybrid pattern greatly improves the accuracies of the \texttt{TW}-only sparse models. 
We propose a pruning algorithm that iteratively shapes the weight matrix to meet our hybrid sparsity pattern constraint. Critically, our pruning algorithm dynamically allocates the sparsity budget to each layer to exploit the inherently uneven sparsity distribution across layers.
 
To maximize the algorithmic benefits of \texttt{TW}, we provide an efficient software implementation on commodity GPU hardware. 
Two key roadblocks arise as a result of the \texttt{TW} sparsity. First, \texttt{TW} naturally introduces frequently uncoalesced memory accesses due to the pruning pattern. Second, different tiles in \texttt{TW} could have different compute demands due to the different pruning degrees across tiles, which leads to load imbalance and GPU resource under-utilization. We address these challenges through a combination of intelligent data layout and concurrency/batching optimizations.
\proj{} achieves an average of $1.95\times$ ($2.86\times$) latency speedup on the tensor core (CUDA core) with only negligible accuracy loss (1\%-3\%). 

\vspace*{0.2cm}
The contribution of our work is as follows:
\begin{itemize}[leftmargin=*]
    \item We propose a tiling-based sparsity pattern to balance the model accuracy and execution efficiency on the existing dense accelerator. The tile sparsity can be combined with the existing fine-grained pattern to minimize the accuracy loss.
    \item We propose a multi-stage pruning algorithm that gradually shapes the weight matrix to our proposed pattern and dynamically allocates the sparsity budget at the layer level to overcome the uneven distribution of sparsity.
    \item We provide an efficient implementation of tile sparsity on commodity GPUs equipped with tenor core, and demonstrate significant speedups on state-of-the-art DNN models.
\end{itemize}

We organize the paper as follows. \Sec{sec:background} describes the background and \Sec{sec:analysis} provides the motivation. \Sec{sec:tile_sparsity} describes the overview of \proj.
We explain its pruning algorithm and tensor core implementation in \Sec{sec:pruning_algorithm} and \Sec{sec:implementation}, respectively.
We evaluate \proj{} against other patterns in \Sec{sec:evaluation}, discuss the related work in \Sec{sec:related_work}, and conclude in \Sec{sec:conclude}.
\section{Background}
\label{sec:background}
This section provides the relevant background on the different deep neural networks that we evaluate in this paper.
We then summarize the recent efforts for reducing the execution latency of those models, which include building specialized hardware accelerators and applying algorithmic pruning optimization to reduce the size and computation cost of DNN models.

\subsection{Deep Neural Network Model} 
\label{background:bert}

DNN models have recently achieved state-of-the-art results in many important domains,  such as convolution neural network~\cite{lecun1995convolutional} (CNN) in the computer vision domain, and long short term memory~\cite{hochreiter1997long} (LSTM, most popular RNN) and BERT~\cite{devlin2018bert} in the natural language processing domain.  
CNN and LSTM are relatively well-studied models whose details are shown in \Fig{fig:bert}. 
We refer the readers to the prior literatures~\cite{chen2016eyeriss, lstmbbs} for more explanations of those models.

BERT is a representative Transformer~\cite{vaswani2017attention}-based model and has outperformed LSTM in NLP domains.
\Fig{fig:bert} shows the structure of a Transformer layer constructing the BERT model. 
The Transformer applies multi-head attention mechanisms, which stands for several groups of independent attentions enabling them to deal with different aspects of information \cite{clark2019does}. 
The BERT model is extremely large and computational expensive.
With adjustable depth and width, there are two popular BERT versions: BERT-large with 24 layers and 16 heads, BERT-base with 12 layers and 12 heads. Without loss of generality, the exploration of this work is built on BERT-base and we refer BERT-base as BERT in the following.

\begin{figure}[t]
    %\vspace*{-3mm}
    \begin{center}
    \includegraphics[width=\columnwidth]{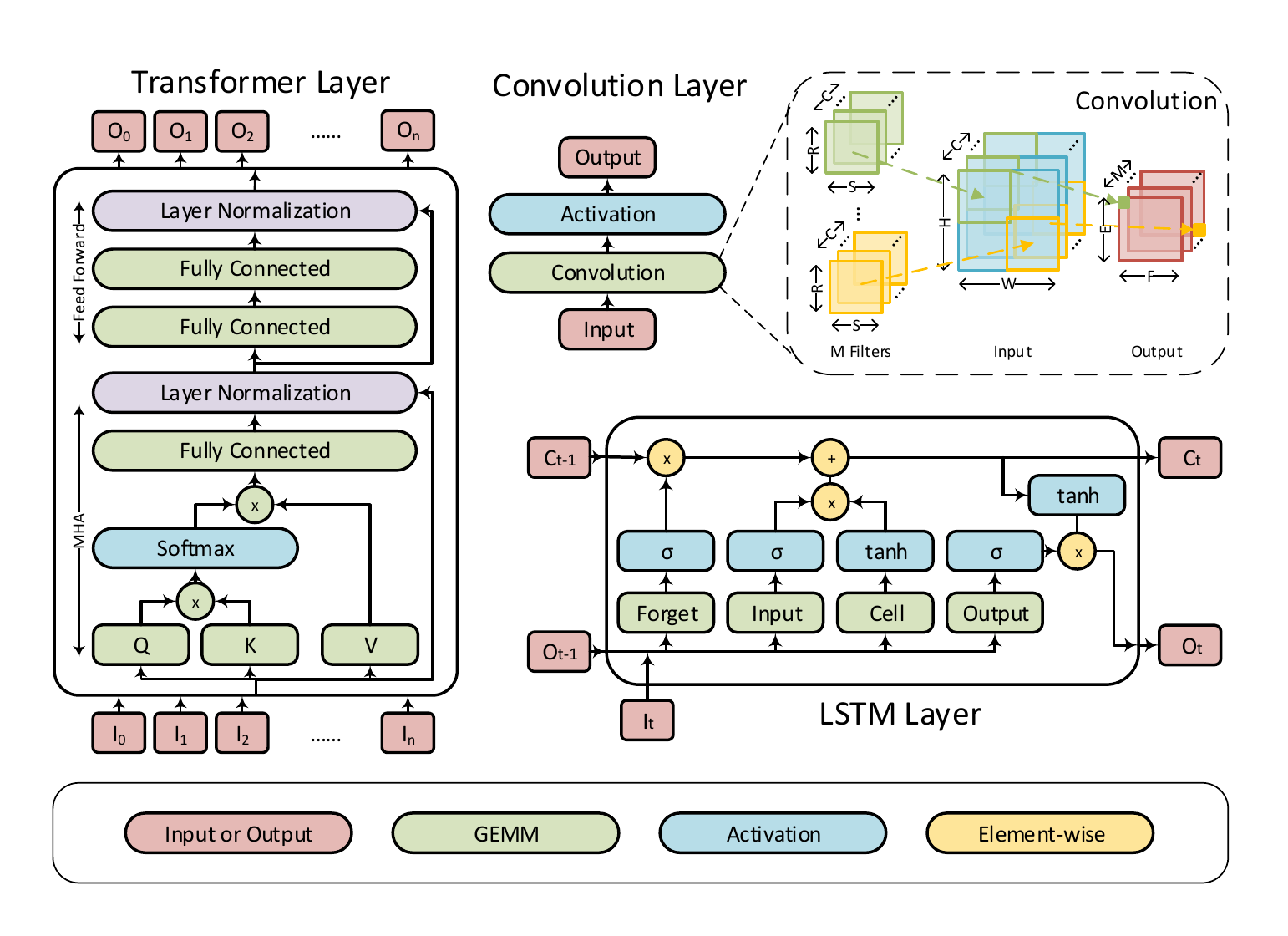}
    %\vspace*{-5mm}
    \caption{The Transformer layer of BERT, convolution layer, and LSTM layer with different kinds of computations.}
    \label{fig:bert}
    \end{center}
    %\vspace*{-4mm}
\end{figure}

\subsection{Hardware Acceleration}\label{background:execution}

We explain the computation characteristics of these models and common optimizations to reduce their execution latency.

\paragraph{Dense Model.}
General matrix multiplication (GEMM) is a key computation in the original dense DNN models, as indicated by the light green blocks in \Fig{fig:bert}. 
The fully connected layer and LSTM layer are native GEMM operations while the convolutional layer can be converted to GEMM through the \benchmark{img2col} transformation.
The attention heads in BERT can also be computed with GEMM operations and the computation of multiple heads could be combined to one large GEMM. 

\noindent\textbf{GEMM Accelerator.}
To reduce the model execution latency, NVIDIA adds tensor core on the GPU in Volta architecture, which runs a fixed size matrix multiplication. Tensor core is essentially an accelerator for the GEMM.
Another example for GEMM accelerator is TPU~\cite{jouppi2017datacenter} which is based on a $128 \times 128$ systolic array.
The cuDNN~\cite{chetlur2014cudnn} library implements different DNN layers for efficient execution on GPU, where the GEMM computation can use the closed-source cuBLAS library~\cite{nvidia2019toolkit} or open-sourced CUTLASS library~\cite{cutlass2019}.

\paragraph{Sparse Model.}
Recently, researchers start to apply pruning~\cite{lecun1990optimal, han2015learning} to DNN models, which exploit the inherent redundancy in the model to transform the original, dense model to a sparse model. In the end, the sparse model has fewer parameters and, theoretically, less computation cost.
Executing sparse models relies on sparse matrix representation such as compressed sparse row (CSR) and sparse GEMM operations, which are supported on GPU by cuSparse~\cite{nvidia2019toolkit} library.
However, as the GPU is originally designed for dense operations, the speedup of sparse model over the dense model is usually negative unless the sparsity ratio is very large (over 95\% reported by prior work~\cite{wen2016learning}). 
As such, researchers begin to put various shape constraints on the pruning pattern and also propose to transform the existing architecture to execute those sparse models.
 For example, the recent work proposes new sparsity patterns that need to modify the existing dense GEMM accelerator like tensor core~\cite{zhu2019sparse} and TPU~\cite{columncombining}.

Different from the prior microarchitecture-centric work, we propose a software-only acceleration of sparse DNN models on the dense GEMM accelerator like tensor core.
We exploit the tile execution of GEMM computation and propose a tiling-friendly, \benchmark{Tile-Wise} sparsity pattern to balance the model accuracy and compatibility for the dense GEMM accelerator. 

\begin{figure}[t]
    \begin{center}
        % \vspace{-3mm}
    \includegraphics[width=1\columnwidth]{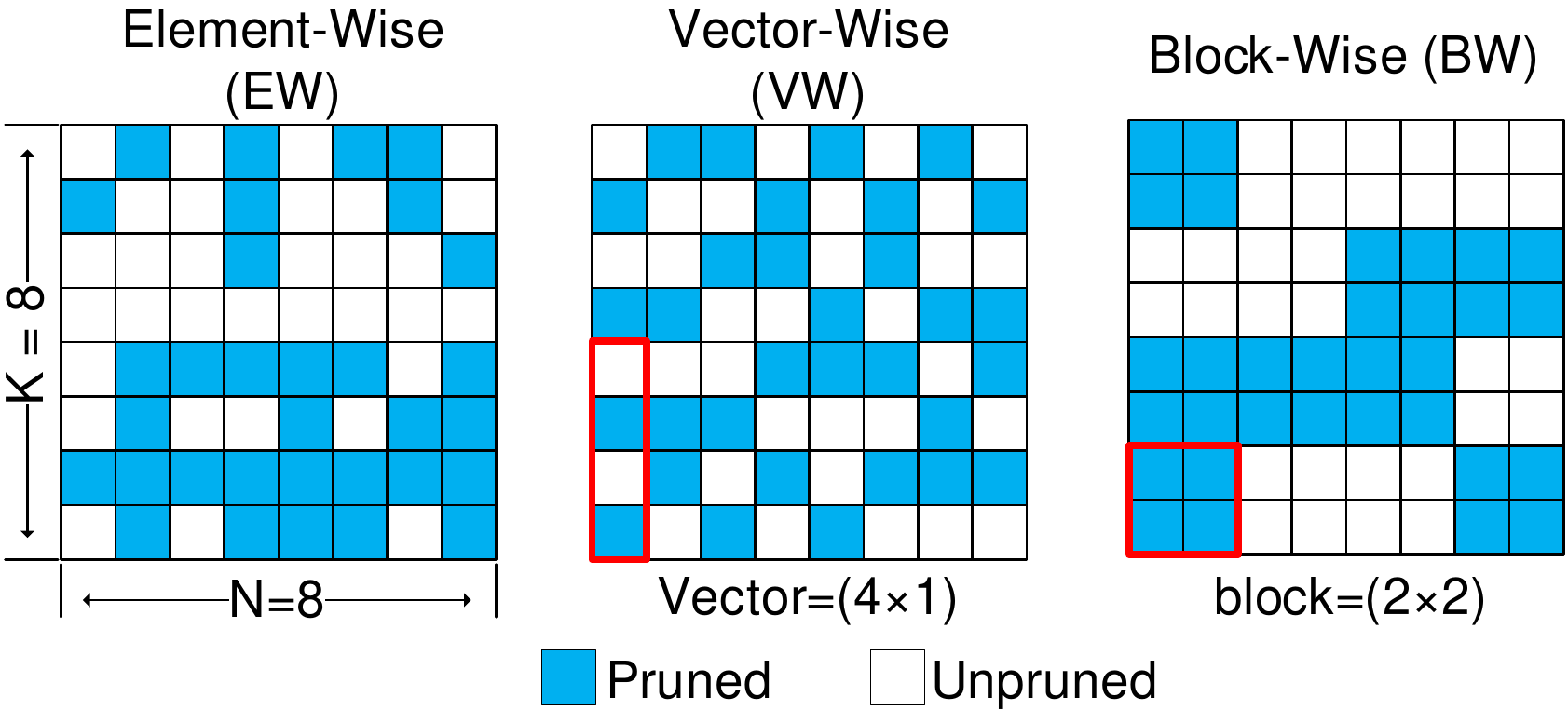}
    \caption{Comparison of three patterns with 50\% sparsity: element-wise (\benchmark{EW}) pattern prunes individual elements, vector-wise (\benchmark{VW}) pattern prunes half elements in a 4-element vector, and block-wise (\benchmark{BW}) pattern prunes an entire $2\times 2$ block.}
    \label{fig:pruning_patterns}
    \end{center}
    %\vspace*{-.4cm}
\end{figure}

\vspace*{0.2cm}
\section{Pruning Efficiency Analysis} 
\label{sec:analysis}

In this section, we study the impact of different DNN pruning algorithms on the model execution efficiency.
Network pruning can remove the excessive weights in DNN models and therefore the resulted sparse model has less weight size and computation cost.
However, our analysis shows that the sparse models generated by existing pruning algorithms cannot achieve meaningful speedup on the dense architectures.

\subsection{Sparsity Pattern}

There are two major components in the algorithms for pruning deep neural networks. The first component is how to evaluate the importance of individual weight element (i.e., importance score). And the second component is the sparsity pattern that is removed altogether.
In this part, we mainly focus on the study of different sparsity patterns and we explain the details of importance score calculation in \Sec{sec:pruning_algorithm}.

\Fig{fig:pruning_patterns} illustrates different sparsity patterns.
The first sparsity pattern, called \benchmark{element-wise (EW)}, removes the individual weight element solely by its importance score rank. 
For instance, prior work~\cite{han2015learning} proposes to remove weight elements with small magnitude.
This approach imposes no constraints on the sparsity pattern and could remove most of weights among all pruning methods.
Thus, it is also called unstructured pruning.
However, the randomly distributed non-zero weights lead to substantial irregular memory accesses, which imposes great challenges for efficient hardware execution.
As such, researchers propose other two more structured pruning methods.

The second sparsity pattern shown in the middle of \Fig{fig:pruning_patterns}, called \benchmark{vector-wise (VW)}~\cite{zhu2019sparse,yao2019balanced}, divides a column in the weight matrix to multiple vectors.
Within each vector, it prunes a fixed portion of elements by the rank of their importance scores.
This approach preserves the randomness within each vector for model accuracy.
Meanwhile, it also maintains the regular structure  for efficient execution, where different vectors have the same number of non-zero weight elements.
The third pattern, called \benchmark{block-wise (BW)}~\cite{narang2017block}, divides the weight matrix to small blocks, and treats a block as the basic pruning unit. 
In other words, the \benchmark{EW} sparsity pattern is a special case of  the \benchmark{BW} sparsity pattern, which expands a $1\times 1$ block to an $n \times n$ block.
The structural sparsity pattern \benchmark{BW} leads to the efficient execution of sparse models.

	\begin{figure}[t]
    \begin{subfigure}{0.48\columnwidth}
	% \vspace*{-.2cm}
	\includegraphics[trim=0 0 0 0, clip, width=.9\linewidth]{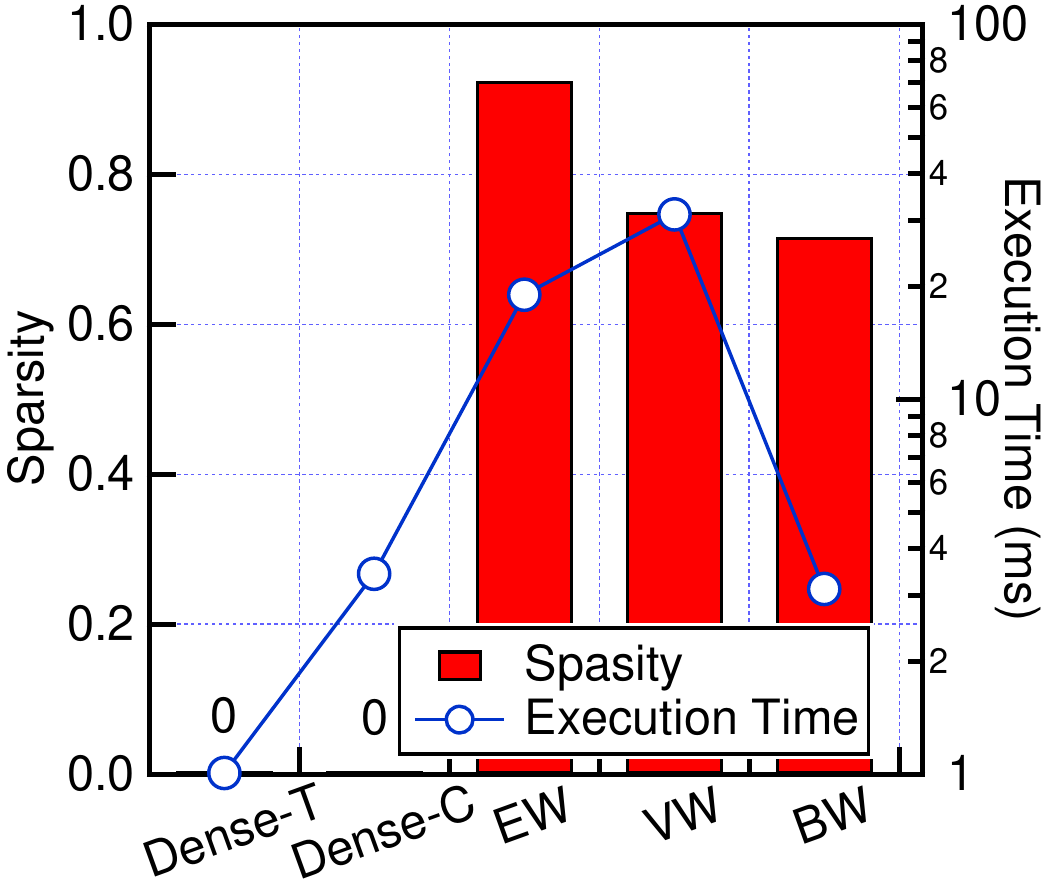}
	    %\vspace*{-.2cm}
    \caption{VGG.}
    \label{fig:sparsity_vgg}
    \end{subfigure}~
    \begin{subfigure}{0.48\columnwidth}
    % \vspace*{-.2cm}
	\includegraphics[trim=0 0 0 0, clip,  width=.9\linewidth]{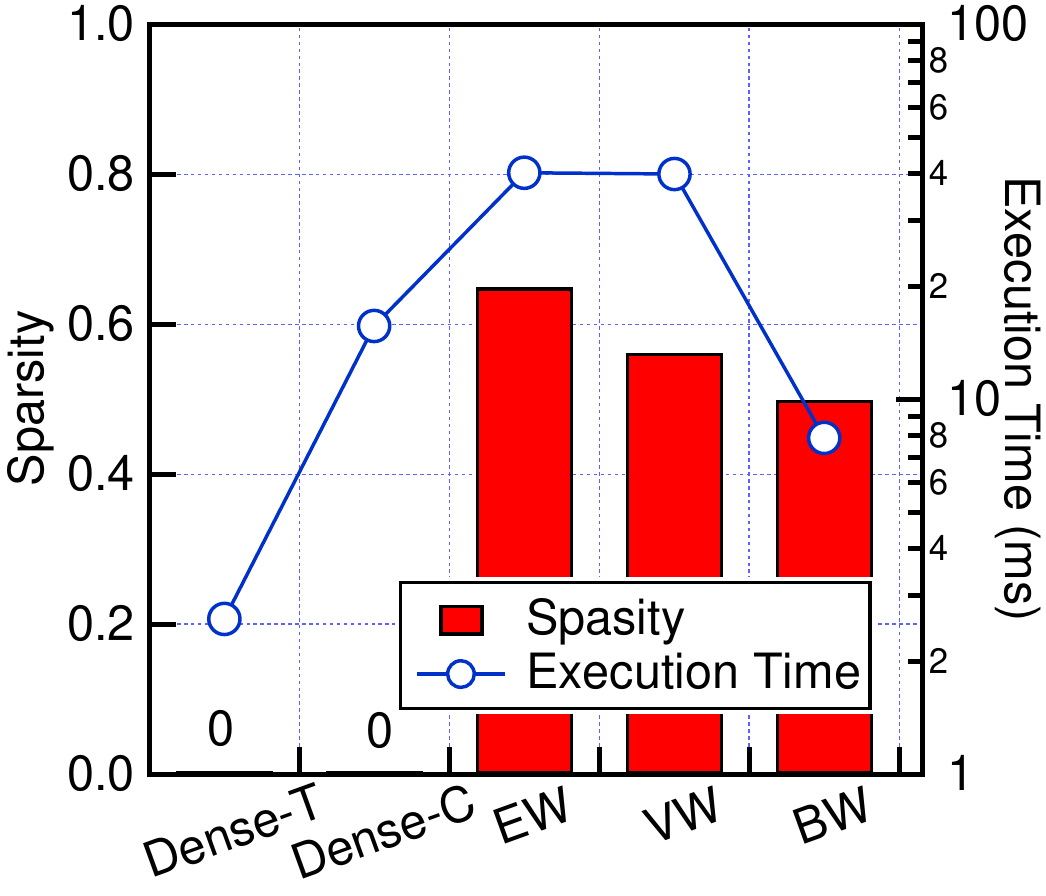}
	    %\vspace*{-.2cm}
    \caption{BERT.}
    \label{fig:sparsity_bert}
    \label{fig:pattern_b}
    \end{subfigure}
	%   \vspace*{-.2cm}
	\caption{Sparsity and execution time comparison between dense and various sparse models (VGG and \hlc{BERT}). \hlc{The accuracy of various sparse models is 1\% lower than the dense model.}}
	\label{fig:sparsity_compare}
	    \vspace*{-.4cm}
	\end{figure}
	
	\begin{figure*}[t]
    \begin{center}
    % \vspace*{-.3cm}
    \includegraphics[width=\linewidth]{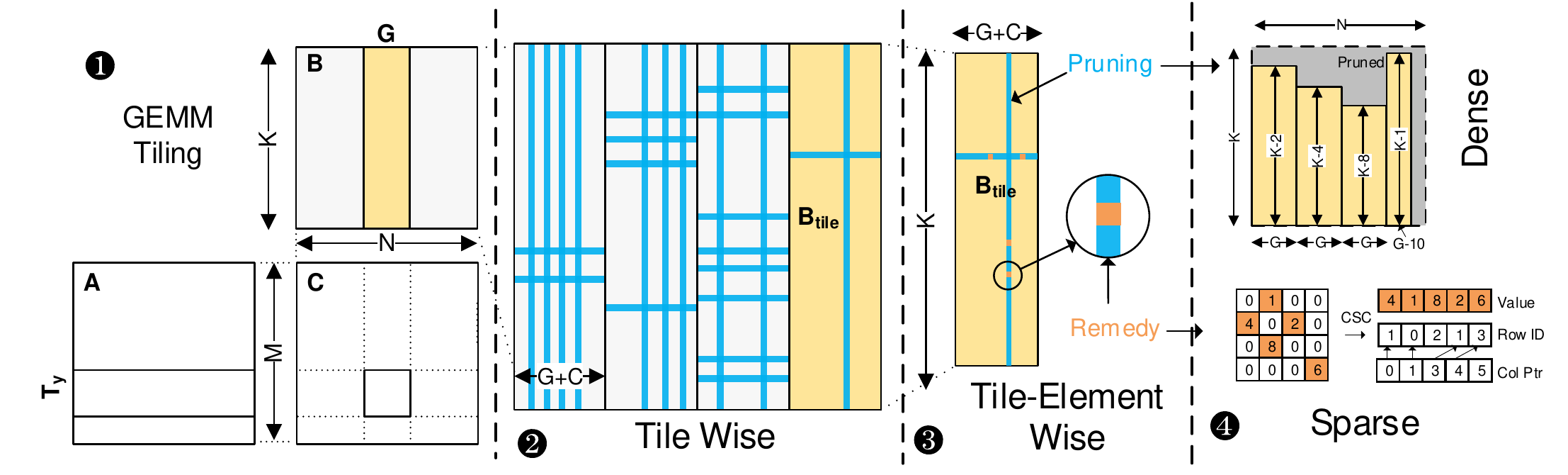}
    %\vspace*{-.2cm}
    \caption{
   The overview of \benchmark{Tile-wise (TW)} sparsity pattern that exploits the tiled dense matrix multiplication (GEMM) to maintain the GEMM-compatible execution. (1) The output matrix tiling based GEMM execution. (2) Our \benchmark{TW} sparsity pattern essentially performs the regular row and column pruning on each matrix tile. (3) The hybrid tile-element-wise (\benchmark{TEW}) sparsity adds back the individual important elements in the pruned row/column to restore the accuracy of sparse \benchmark{TW} models. (4) The execution of \benchmark{TW} can be converted to multiple small dense GEMMs, and the execution of \benchmark{TEW} can use extra sparse GEMM.
    }
    \label{fig:tw}
    \end{center}
    \vspace*{-0.2cm}
\end{figure*}

\subsection{Execution Efficiency Analysis}

We first use two popular deep neural network models to evaluate the execution efficiency of sparse models generated by the aforementioned three sparsity patterns.
Our experimental results show that although these pruning approaches can lead to sparse models with a large volume of sparsity (i.e., zero weight elements), they fail to achieve a meaningful speedup compared to the unmodified dense model on the existing GPUs. 
Moreover, the emergence of dense matrix multiplication accelerators such as tensor core further widens their performance gap.
 
We first study a CNN model VGG-16~\cite{simonyan2015very} with 13 convolutional layers and 3 fully connected layers. 
We evaluate it on the image classification task using the ImageNet dataset~\cite{krizhevsky2017imagenet}.
The second studied model is BERT (base) in \Sec{background:bert}, which is a Transformer-based model with 12 encoding layers.
We evaluate the BERT model with the sentence classification task on the MNLI dataset~\cite{wang2019glue}.
We use the vector size of 16 for \benchmark{VW} and the block size of $32\times32$ for \benchmark{BW} as suggested in their original papers~\cite{zhu2019sparse,narang2017block}.
We prune both models with the three different sparsity patterns and keep the accuracy drop of each model within 1\% of its unmodified dense version.

We perform the efficiency analysis on a V100 GPU~\cite{volta2017whitepaper} with CUDA 10.1~\cite{nvidia2019toolkit}. 
Besides the CUDA cores, the V100 GPU also integrates the specialized tensor core for the acceleration of dense matrix multiplication.
This GPU has a peak throughput of 15.7 TFLOPS (floating point operation per second) and 125 TFLOPS, for the CUDA cores and tensor cores, respectively. 
We evaluate the performance of dense model with the cuDNN library on the CUDA core and tensor core separately.
We execute the sparse model of \benchmark{EW} and \benchmark{VW} with the cuSparse~\cite{nvidia2019toolkit} library, which executes only on the CUDA cores.
We execute the sparse model of \benchmark{BW} with the BlockSparse~\cite{blocksparse2020} library, which leverages the tensor cores owing to its regularity.

\Fig{fig:sparsity_compare} compares the sparsity and performance between the dense and sparse models with various patterns. 
All patterns achieve over 50\% sparsity with the greatest by \benchmark{EW}.
The performance of \benchmark{EW} and \benchmark{VW} is slower than their dense version on the CUDA core (Dense-C). 
In addition, the performance gap between the dense model and sparse model exacerbates when the dense matrix accelerator tensor core is used (Dense-T).
Prior work~\cite{zhu2019sparse} reports a $1.5\times$ speedup using the \benchmark{VW} pattern, which requires non-negligible modifications of the tensor core.
\benchmark{BW} achieves the best performance among all sparse models as it runs on the tensor core.
However, its performance is still $3\times$ slower than the dense model on tensor core.

  In summary, the existing pruning approaches generate sparse models that are inefficient on the existing hardware.
As such, we need a sparsity pattern that can match the existing hardware features while maintaining the fine granularity, which is critical for achieving the high model accuracy.

\vspace*{0.2cm}
\section{Tile Sparsity}
\label{sec:tile_sparsity}

In this section, we present the details of our proposed tile sparsity pattern.
Our approach leverages the tiled execution of matrix multiplication, which is originally designed for exploiting the parallel computation resources.
 The proposed tile sparsity pattern introduces irregularity at the global matrix level, but maintains the regularity of individual matrix tile.
 As such, it can balance the DNN model accuracy and compatibility for dense matrix accelerator, e.g., tensor core.
We also show that the tile sparsity pattern can be overlaid with the most fine-grained element-wise pattern to increase the sparsity of pruned models and reduce their accuracy loss.

\subsection{Tiling and Pruning Co-design}
\label{subsec:tw}

As \Sec{sec:background} explains, the dominant computation in deep neural network models is the general matrix multiplication (GEMM).
In this subsection, we first present the details of tiled matrix multiplication.
We then propose to co-design the matrix tiling and deep neural network pruning, which leads to the \benchmark{tile-wise (TW)} sparsity pattern.
We explain how \benchmark{TW} maintains the compatibility on the dense GEMM accelerator and the composability with the fine-grained sparsity pattern.

\Fig{fig:tw} \circled{1} shows one level tiling of matrix multiplication on the GPU.
The GEMM computes $C = A \times B$ with input matrix $A\ (M\times K)$, weight matrix $B\ (K \times N)$, and output matrix $C\ (M \times N)$. 
Since modern high-performant microprocessors mostly adopt the manycore architecture, the tiled execution of output matrix $C$ breaks the entire GEMM computation into multiple ones such that they can run on multiple cores for the parallel execution.
Specifically, each core (or streaming multi-processor, SM in NVIDIA GPU) computes one tile with size of $T_y \times G$.
Consequently, the core only loads $T_y$ rows of input matrix $A$ and $G$ columns of weight matrix $B$ (called $B_{tile}$).

With the output matrix tile size of $T_y \times G$, the $K\times N$ weight matrix $B$ is divided to $\lceil \frac{N}{G} \rceil$ $B_{tile}$.
The key idea of our \benchmark{tile-wise} pattern is to prune each $B_{tile}$ with the regular row pruning and column pruning. 
As shown in \Fig{fig:tw} \circled{2}, the row pruning treats an entire row of each weight tile $B_{tile}$ as the basic pruning unit, which leads to the reduced K-dimension size (i.e., height) of each $B_{tile}$.
We prune each $B_{tile}$ with different number of rows determined by the pruning algorithm that we describe later. 
The difference across different tiles maintains the irregularity of sparsity that is required by model accuracy. 
E.g., the heights of four weight matrix tiles in \Fig{fig:tw} \circled{2} are $K-2$, $K-4$, $K-8$, and $K-1$ respectively after the row pruning.

Besides the row pruning, we also perform the column pruning for the weight matrix tile $B_{tile}$, which reduces its N-dimension size.
Our approach prunes different number of columns ($C$) in each weight matrix tile for better irregularity.
The combined row and column pruning alleviate the constraint on the sparsity pattern and therefore allow more weight elements to be pruned.
In specific, we perform column pruning before row pruning for maximizing the execution efficiency, which we explain later.

\paragraph{Pattern Overlay.}
Since the \benchmark{TW} still enforces a particular pruning pattern, important weight elements could be removed, which may lead to accuracy loss.
We propose to overlay \benchmark{TW} and \benchmark{EW} to mitigate the accuracy loss.
\Fig{fig:tw} \circled{3} illustrates the resulted hybrid pattern \benchmark{tile-element-wise (TEW)}. 
In order to prune $\alpha$ percent of weights, the \benchmark{TEW} first prunes $\alpha+\delta$ percent of weights with only \benchmark{TW}, and then \textbf{restores} $\delta$ percent of the weight elements with the highest importance scores.

\paragraph{Pruning Order.}
To improve the execution efficiency of  the proposed sparsity pattern, we first perform the column pruning and then re-organize the weight matrix tiles for row pruning.
We use the example in \Fig{fig:tw} \circled{2} to illustrate its advantages.
With the column pruning, the four tiles are pruned with 4, 3, 2, and 1 columns, respectively.
For the row pruning inside each tile, we re-organize the four tiles with $G+4$, $G+3$, $G+2$, and $G-9$ columns.
After pruning (\Fig{fig:tw} \circled{4}), the N-dimension sizes of the four tiles are $G$, $G$, $G$, and $G-10$, respectively.
The first three tiles have the same number of columns such that their execution can be batched for better performance.
For the hybrid \benchmark{TEW} pattern, each tile stores the \benchmark{EW} pattern with the compressed sparse column (CSC) format.
We leverage linear property of matrix multiplication to execute the \benchmark{TW} and \benchmark{EW} separately.
We explain the execution details in \Sec{sec:implementation}.

\subsection{Comparison with Other Sparsity Patterns}
\label{sparsity:comparison}

In this subsection, we demonstrate that the \benchmark{TW} is not only friendly for the hardware execution, and also preserves the model accuracy.
In specific, we compare the irregularity, which determines the pruned model accuracy under the same sparsity percentage,  of \benchmark{EW}, \benchmark{TW}, \benchmark{BW}, and \benchmark{VW}.
We use the \benchmark{EW} as the baseline because it has the highest degree of irregularity and therefore the best accuracy.

	\begin{figure}[t]
	% \vspace*{-.2cm}
    \begin{center}
    \minipage{0.48\linewidth}
     \includegraphics[width=\linewidth]{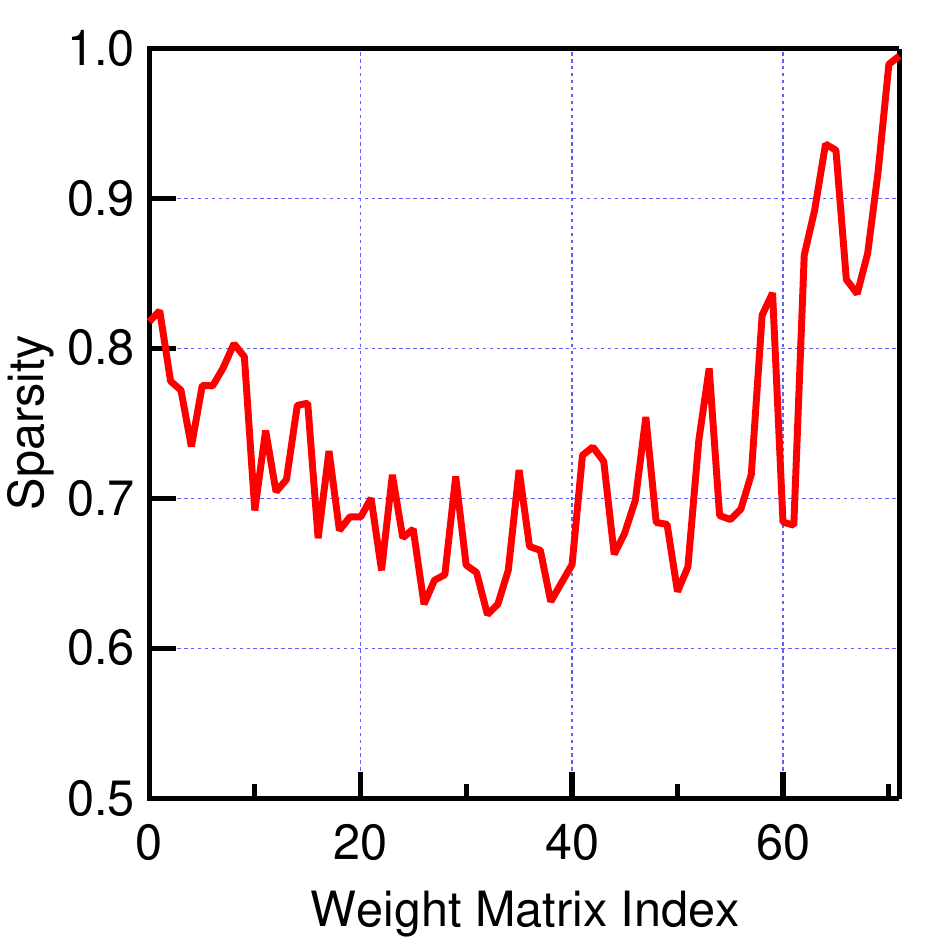}
    \caption{\small \hlc{The per-layer sparsity when pruning BERT using the \mbox{\benchmark{EW}} pattern with 75\% overall sparsity.}}
    \label{fig:ew_res}
    \endminipage\hfill%
    \minipage{0.48\linewidth}
       \includegraphics[width=\linewidth]{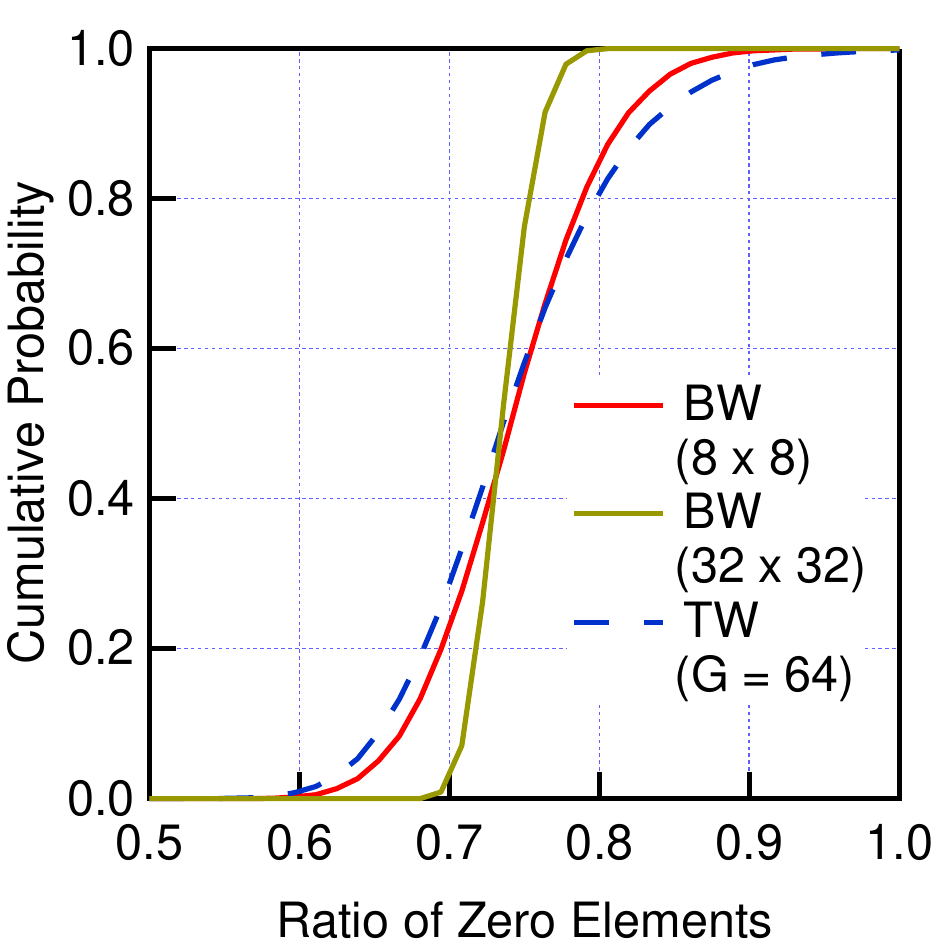}
    \caption{\small Cumulative probability distribution of zero weight elements with \benchmark{BW} and \benchmark{TW} patterns.}
    \label{fig:zero_density}
    \endminipage
    \vspace*{-0.2cm}
	\end{center}
    \end{figure}

\paragraph{Against \benchmark{VW}.}
First, we find that there exists uneven distribution of sparsity in different weight matrices, which makes \benchmark{TW} a more suitable pattern than \benchmark{VW}.
We illustrate this point by showing the sparsity distribution of 72 weight matrices in BERT, which has 12 layers and each layer has 6 weight matrices (4 for the self attention and 2 for FC layers).
We apply the \benchmark{EW} pattern to prune 75\% weights.
In specific, the importance score of all elements in the 72 weight matrices are calculated and globally ranked for the element-wise pruning.
As \Fig{fig:ew_res} shows, the final pruned weight matrices have different degrees of sparsity although the averaged sparsity for all weights is 75\%.
This suggests the uneven distribution of sparsity in the DNN model.
\benchmark{VW} splits every column to a certain number of groups and all the groups have the same sparsity by pruning the same number of elements.
In contrast,  \benchmark{TW} can maintain the uneven sparsity distribution by globally ranking the matrix tiles.

\paragraph{Against \benchmark{BW}.}
Compared to the \benchmark{BW}~\cite{narang2017block}, \benchmark{TW} can remove more weights owing to its fewer constraints on the pruning shape.
To illustrate this point, we characterize the number of zero elements in different pruning shapes on the BERT model with 75\% \benchmark{EW} sparsity.
\Fig{fig:zero_density} compares the number of zero elements in a block of $8\times 8$, $32\times32$ for the \benchmark{BW}, as well as in a row vector of  64 elements for the \benchmark{TW} with $G=64$.
Both with 64 elements, \benchmark{TW} captures more zeros elements than \benchmark{BW}.
 Meanwhile, prior work reports that \benchmark{BW} requires a pruning unit of $32\times 32$ for maintaining high performance~\cite{child2019generating}, which captures even fewer zero elements as \Fig{fig:zero_density} shows.
 In contrast, \benchmark{TW} with $G=64$ is sufficient for achieving significant speedups as we show later. 
 
 In summary, we conclude with the irregularity relationship of \benchmark{EW} $>$ \benchmark{TW}  $>$ \benchmark{VW} $\approx$  \benchmark{BW}. 
Owing to the existence of globally uneven sparsity distribution, \benchmark{TW} leads to a pattern that is closer to \benchmark{EW} than the \benchmark{VW} pattern. 
\benchmark{TW} also removes more weights than \benchmark{BW} owing to its fewer constraints on the pruning shape.
\setlength{\textfloatsep}{5pt}
\begin{algorithm}[t]
    \caption{The multi-stage \benchmark{TW} pruning algorithm.}
    \label{alg:pruning}
    \small
\KwIn{Pre-trained weight matrix, $M_s$, and shape, ($K$, $N$); \\
\hspace*{1cm}Target sparsity, $S$; Variable granularity, $G$;}

\KwOut{Pruned weight matrix, $M_d$}
    $m=M_s$; $s_t=0$; \\
\While{$s_t<S$}{
    $s_t$=GraduallyIncrease($s_t$);\\
    $m =$ $m$ Splited by shape($K$, 1) for Column Pruning;\\
    $tileScore=$ImportanceScore($m$);\\
    $tileScore=$AprioriTuning($tileScore$, $S$);\\
    $threshold=$ Percentile($tileScore$, $s_t$);\\
    \While{each $tile_i \in m$}{
        \If{$tileScore[i] < threshold$}{
            Prune the Column $tile_i$ with shape(K, 1);\\
        }
    }

    $m =$ $m$ Splited by shape(1, $G$) for Row Pruning;\\
    $tileScore=$ImportanceScore($m$);\\
    $threshold=$ Percentile($tileScore$, $s_t$);\\
    \While{each $tile_i \in m$}{
        \If{$tileScore[i] < threshold$}{
            Prune the Row $tile_i$ with shape(1, G);\\
        }
    }
    FineTune($m$);\\
}
$M_d=m$; return $M_d$;
\end{algorithm}

\vspace*{0.2cm}
\section{Tile Sparsity Based Pruning} 
\label{sec:pruning_algorithm}

This section explains our multi-stage pruning algorithm for leveraging the proposed \benchmark{TW} sparsity pattern. 
Algorithm~\ref{alg:pruning} describes the algorithm, which we explain in details as follows.

\paragraph{Overview.} 
We adopt the multi-stage pruning algorithm that gradually prunes the pre-trained dense model to reach a target sparsity.
Each stage consists of a pruning and fine-tuning step, where the algorithm first prunes the model with a small sparsity target and then fine-tunes the pruned model to restore the model accuracy.
The pruning stage is repeated until the model reaches the target sparsity.
Prior work points out that the multi-stage pruning improves the model accuracy than the single-stage pruning~\cite{han2015learning}.
At each stage, the algorithm calculates the importance score of each tile. It then performs the column and row pruning according to the rank of importance score.
We also perform the apriori tuning that borrows the information of \benchmark{EW} pruning to reduce the accuracy loss.
Each iteration from line 3 to 21 in Algorithm~\ref{alg:pruning} is a complete pruning-tuning stage, where line 3 increments the target in the current stage.

\begin{algorithm}[t]
    \caption{Apriori tuning.}
    \label{alg:optimization}
    %\begin{algorithmic}[1]%一行一个标行号
\small
\KwIn{EW pruned results, $EW$; Top-n and Last-n, $n_1$, $n_2$;\\
\hspace*{.9cm} Tile importance score, $tileScore$; Target sparsity, $S$;}

\KwOut{Tile importance score, $tileScore$;}
    $tileSparsity=EW[S]$; \\
    $topNTiles=$GetTopNTiles($tileScore$, $n1$);\\
    $tileScore=$SetZero($tileScore$, $topNTiles$) \\
    $lastNTiles=$GetLastNTiles($tileScore$, $n2$);\\
    $tileScore=$SetInf($tileScore$, $lastNTiles$);    return $tileScore$;
\end{algorithm}

\begin{figure*}[t]
    %\vspace*{-.4cm}
    \hspace{-2mm}
    %\begin{center}
    \includegraphics[width=2.1\columnwidth]{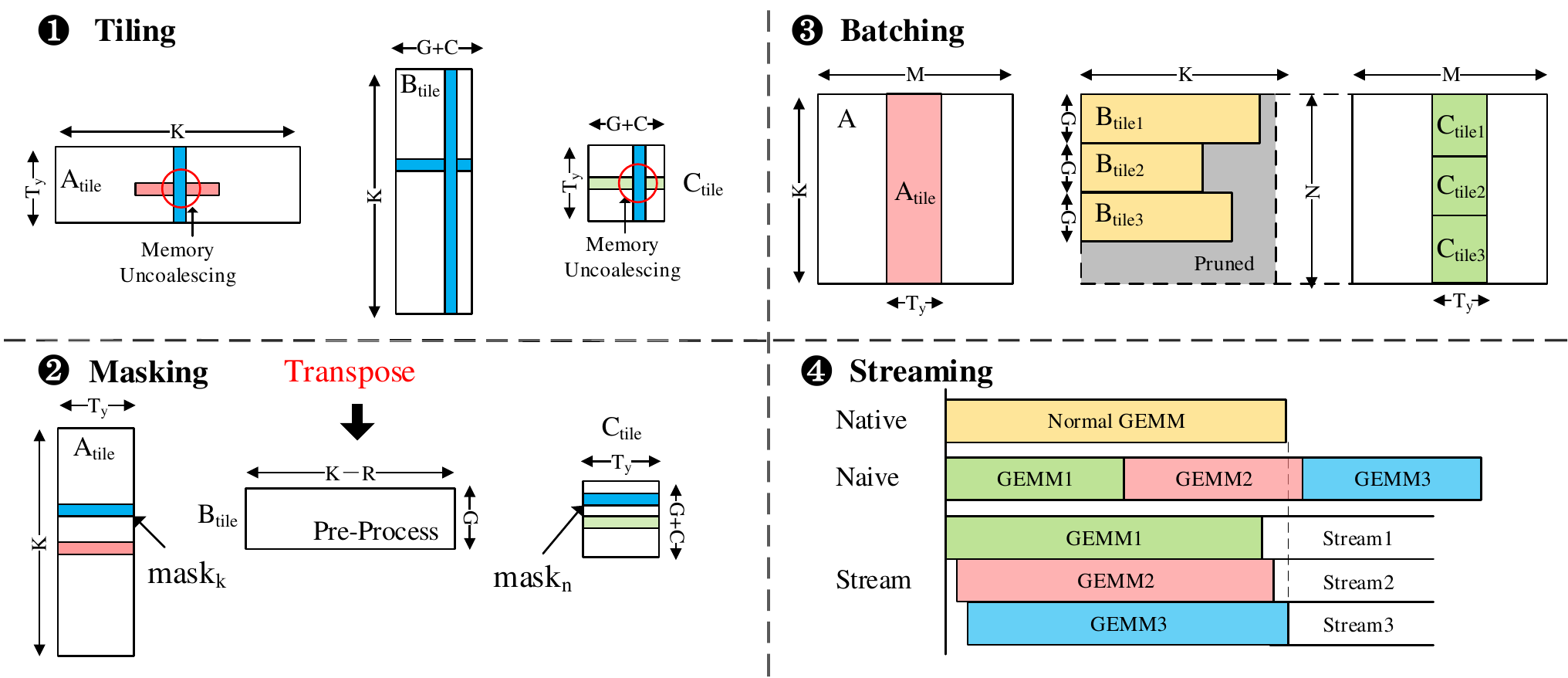}
    \caption{
    Our implementation performs a series of steps to optimize the execution efficiency of \benchmark{TW}-based sparse models. 
    (1) Naive tiling leads to uncoalesced memory accesses and therefore performance loss. 
    (2) We transpose the matrix for memory access coalescing and use masks for efficient processes.
	(3) We convert the execution of \mbox{\benchmark{TW}} to batched GEMM that lets us reuse the existing tensor core-based GEMM kernels. 
	(4) We improve the execution efficiency by leveraging the stream concurrency.}
    \label{fig:streamgemm}
    %\vspace*{-.3cm}
    %\end{center}
    \end{figure*}
    
\paragraph{Importance Score.}
How to compute the importance score is an active research topic~\cite{han2015learning,luo2017thinet,yu2018nisp,Zhang2018StructADMMAS, molchanov2016pruning, molchanov2019importance, qiu2019adversarial}.
The most intuitive approach~\cite{han2015learning} is to use the weight's absolute value.
We use a more accurate approach~\cite{molchanov2019importance} that uses the incurred error by removing a parameter as its importance score.
\hlc{Although not the focus of our work, we find that this approach leads to a better accuracy under the same sparsity on complex models like BERT, and can reduce the fine-tune time for other models like NMT and VGG in our experiments.}

Let $ L $ be the loss function, $w$ be the targeted weight variable, $w = w_i$ means the original value, and $w = 0$ means that we prune $w$. The importance score for $w$ is the difference of loss function as Equ.~\eqref{eq1} shows.
However, the exact computation is expensive because $M$ parameters require evaluating $M$ versions of the network, one for each parameter.
We avoid evaluating $M$ different networks by approximating $\Delta L(w)$ in the vicinity of $w$ by its first-order Taylor expansion in Equ.~\eqref{eq2} where $R_1$  is the remaining term of first-order Taylor expansion, as suggested by the prior art~\cite{molchanov2019importance}.
The approximated importance score in Equ.~\eqref{eq3} is essentially the product between weight value $w_i$ and weight gradient $\partial L(w_i)/\partial w $, both of which already exist in the training stage and therefore are easy to derive.
\begin{align} 
{ \Delta L(w)}  &= \sqrt{({ L(w = w_i) - L(w = 0) })^2}  \label{eq1}\\
L(w = 0) &=  L(w_i) + \frac{\partial L(w_i)} {\partial w} * w_i + R_1 ( w = 0) \label{eq2} \\
{ \Delta L(w)}  &\approx \sqrt{({ \frac{\partial L(w_i)}{\partial w} * w_i })^2} \label{eq3}
\end{align}

\paragraph{Pattern Pruning.}
As explained in \Sec{sec:tile_sparsity}, the \proj{} pattern requires the column pruning before the row pruning.
We first break the weight matrix into column-based tiles (line 4) and then evaluate the importance score of each tile.
We then determine the threshold for column pruning based on the sparsity target in the current stage (line 5).
Line 6 applies the apriori tuning that we explain later, and line 8-12 remove the column tiles.
Afterwards, we reorganize the column-pruned matrix to tiles (line 13).
Line 14-20 perform the row-pruning, which is similar to the column pruning.
After the column and row pruning, the weight matrix becomes compatible with \proj{}.
In our algorithm, we design the tiling granularity $G$ as a tunable hyper-parameter, through which we explore the trade-off between the accuracy and performance of the sparse model.

\paragraph{Global Weight Pruning.}
As shown in \Sec{sparsity:comparison}, there exists an uneven distribution of weight sparsity in different layers of a DNN model, which we use a global weight pruning to exploit.
The codes in line 7 and line 15  sort the scores for all tiles in the column and row pruning, respectively.
 The codes in line 8-12 and line 16-20 prune the tiles from all layers in the DNN model according to their importance rank.

\paragraph{Apriori Tuning.} We use \benchmark{EW} pattern with the target sparsity as an apriori knowledge to better guide our \proj{}-based pruning algorithm because \benchmark{EW} achieves the best model accuracy under the same sparsity. 
In the \benchmark{EW} sparsity results, we observe a strong locality pattern, where more than 10\% tiles (columns) are completely pruned (i.e., 100\% sparsity) when the pruning target sparsity is 75\%. 
We leverage this observation to augment our pruning algorithm with apriori tuning in Line 6 of Algorithm~\ref{alg:pruning}. 
The detailed apriori tuning algorithm is shown in Algorithm~\ref{alg:optimization}. 
First, we get the tile-level sparsity distribution from the \benchmark{EW} results in the target sparsity. We set the top-n maximum sparsity tile score 0, which means high priority to prune. In contrast, we set the last-n tiles a large score and would not be pruned.

\section{Efficient GPU Implementation} \label{sec:implementation}

This section introduces our efficient GPU implementation that unleashes the algorithmic benefits of \texttt{TW}. Exploiting the unique sparsity pattern of \texttt{TW}, we first describe the basic tiling design, followed by three key optimizations that combine intelligent data layout and concurrency/batching optimizations to maximize the efficiency of \texttt{TW} tiling on tensor cores.

The advantage of \proj{} sparsity pattern is that sparse matrix multiplication could be transformed to dense GEMM, which can be effectively accelerated on dense GEMM accelerators such as the tensor core on GPUs (\Sec{sec:tile_sparsity}).
\Fig{fig:streamgemm} shows how we transform sparse matrix multiplication that has the \texttt{TW} sparsity pattern to a dense GEMM, and how it exploits various GPU characteristics to maximize the performance.

\paragraph{Tiling.} We start by tiling matrices as usual. \Fig{fig:streamgemm} \circled{1} illustrates an example, where generating an output tile  $C_{tile}$ requires two input tiles $A_{tile}$ and $B_{tile}$. Each input matrix tile has two mask vectors that indicate which rows and columns in the matrix tile are pruned. In the example of \circled{1}, the blue rows and columns are pruned. We remove the pruned rows and columns in the weight matrix tile $B_{tile}$, which can be done offline before the model inference starts.
The input tile $A_{tile}$ and output tile $C_{tile}$ are stored in the dense format to avoid the preprocess overhead. Their pruned rows/columns are skipped rather than removed.

We modify the dense GEMM kernel such that it skips computing partial sums for pruned elements according to the mask vectors. This reduction of computation is the source of acceleration. Our baseline GEMM implementation is based on the open-sourced CUTLASS~\cite{cutlass2019}, which is a high-performance linear algebra CUDA library. It implements three levels of tiling to maximize the data reuse in the global memory (thread block tile), shared memory (warp tile), and register file (thread fragment).
Meanwhile, it can also leverage the tensor core in the GEMM computation, which we use to accelerate the \proj{}.

However, a naive tiling implementation is inefficient and even causes slowdown compared to the original dense model. In our implementation, we exploit three optimizations that mitigate the inefficiencies and maximize the benefits of \texttt{TW}.

\paragraph{Memory Accesses Coalesce.} Naive tiling leads to frequent uncoalesced memory accesses that are inefficient on the single-instruction-multiple-data based GPUs. \Fig{fig:streamgemm} \circled{1} shows the memory access patterns in the original row-major matrix format.
The pruned blue row in $B_{tile}$ causes the skip of blue column in $A_{tile}$.
Therefore, a continuous access to the $A_{tile}$ (marked by the red row) that is originally coalesced now becomes uncoalesced, which can cause severe performance degradation as uncoalesced memory accesses require multiple memory transactions.
The uncoalesced accesses also exist in the matrix tile $C_{tile}$ (marked by the green row) owing to the pruned blue column in $B_{tile}$.

We propose to store the matrix tiles in their transposed format to optimize their memory access efficiency.
In \Fig{fig:streamgemm} \circled{2} where the three tiles are transposed, the column skipping is converted to the row skipping. 
Thus, it eliminates the uncoalesced accesses and improves the access efficiency.

\begin{figure}[t]
	%\vspace*{-0.3cm}
    %\hspace{-2mm}
    \centering
    \includegraphics[width=.8\columnwidth]{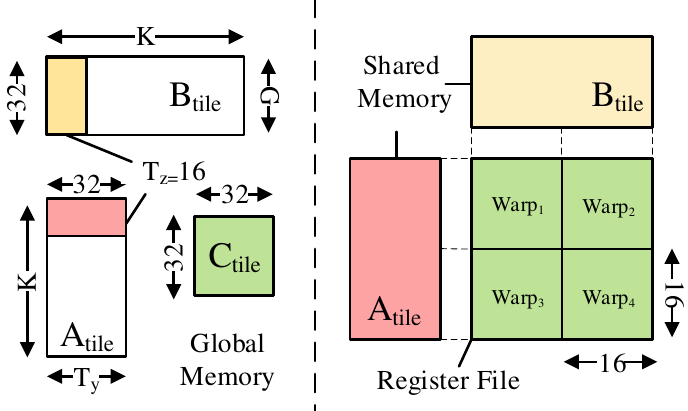}
    \caption{Warp-level GEMM tiling that exploits tensor core.}
    \label{fig:warplevel}
\end{figure}

\paragraph{Load Imbalance Mitigation.}  \texttt{TW} sparsity inherently introduces imbalanced tiles. That is, some tiles will require more computations since fewer rows/columns are pruned; other tiles that have more rows/columns pruned will lead to lower computation. Imbalanced tiles lead to resource under-utilization, and thus affects the overall speedup.

We propose to batch tile computations to improve the utilization. \Fig{fig:streamgemm} \circled{3} shows an example where the weight matrix is decomposed into $\lceil \frac{N}{G} \rceil$ tiles, where $G$ is the \texttt{TW} granularity.
Different $B_{tile}$ are batched together to share the same $A_{tile}$. Batching improves resource utilization as a batched GEMM packs multiple tiles and thus increases the computation.

Another practical benefit of batched-GEMM implementation is that we can reuse existing high-performance tensor core-based GEMM kernels and avoid implementing specialized GEMM kernels, each customized for a particular tile size.
\Fig{fig:warplevel} illustrates the warp-level tiling and Listing~\ref{lst:gemm} shows the kernel implementation that uses tensor core APIs.
We assume that $G=32$, $T_y=32$ and $T_z=16$, which is the minimum tiling granularity as it must be the multiple of $32$ (i.e., warp size). $A_{tile}$ and $B_{tile}$ are stored into the shared memory and $C_{tile}$ to the register file. Then a warp tile will compute the out-product with the tensor core MMA API, which can support the fixed size ($16 \times 16\times 16$) matrix multiplication.

While batching mitigates resource under-utilization, we find that it is possible that the computation of a batch still under-utilizes the GPU resources. We leverage concurrent kernel execution on modern GPUs~\cite{streams2015} to further improve resource utilization. 
In the studied NVIDIA GPU platform, we overlap the computation of different tiles by assigning to different streams, and rely on the underlying scheduler to maximize resource utilization. \Fig{fig:streamgemm} \circled{4} shows an example where naively running different batches could have lower performance than the original unpruned GEMM. Concurrently executing multiple batches with different streams improves performance.

{
\begin{lstlisting}[float=t, language=C++, caption={GEMM kernel on tensor core.}, label={lst:gemm}]
#define G   32
#define T_y 32
#define T_z 16
__global__ void StreamMaskedGEMM(int M, int Pruned_N, int Pruned_K, half *A, half *B, half *C, half alpha, half beta, int *mask_n, int *mask_k){    
    //Allocate C_tile in Register File.
    half C_tile[G * T_y];
    //Allocate A_tile and B_tile in Shared Memory.
    __shared__ half A_tile[T_y * T_z];
    __shared__ half B_tile[G   * T_y];
    for(int k = 0; k < K; k+=T_z){
        //Load A_tile from Global to Shared Memory skipping the pruned row with mask_k.
        Load_A_Tile_with_Mask(A_tile, A, mask_k);  
        //Load B_tile from Global to Shared Memory with Pre-Processed B.
        Load_B_Tile(B_tile, B);
        //Tensor core API: WMMA with fixed 16x16x16 GEMM.
        WMMA::MMA(C_tile, A_tile, B_tile, alpha, beta);
    }
    //Store C_tile from Register File to Global Memory skipping the pruned row with mask_n.
    Store_C_Tile_with_Mask(C, C_tile, mask_n);
}
\end{lstlisting}
}

\paragraph{Kernel Fusion.} Complex DNN models necessarily incorporate non-GEMM computations. For instance, the BERT model spends about {39\%} time on non-GEMM kernels, such as \benchmark{Add-bias} and \benchmark{LayerNormalization}, constituting the ``Amdahl's law bottleneck.''
Meanwhile, our memory coalescing optimization also introduces additional transpose kernels, which can incur a large performance overhead if leave unoptimized. 

As such, we propose to fuse consecutive non-GEMM kernels to improve the performance. Kernel fusion has two advantages. First, we reduce the number of kernels to reduce the launch time. Second, fused kernel reduces access to global memory and shares the register resources. For example, the previous \benchmark{Add-bias} operation can execute with \benchmark{LayerNormalization} when the data is loaded into the register file.
We also modify the memory access behavior in those non-GEMM kernels to reduce the number of transpose kernels.
With this modification, we only need to transpose matrix $A$ in the first layer and transpose matrix $C$ after the last layer, which significantly reduces the transpose overhead.
\hlc{We also apply the kernel fusion optimization to the dense model baselines for the fair comparison, which, for example, reduces the 39\% non-GEMM execution time in BERT to 29\%.}
\section{Evaluation} \label{sec:evaluation}

In this section, we demonstrate that \proj{} is able to maintain the accuracy of sparse DNN models and provide the significant execution speedup over the dense model and other sparsity patterns at the same time. 
We first explain our evaluation methodology with the use of state-of-the-art DNN models on the GPU equipped with tensor cores.
We then study the design space of \proj{} to explore the trade-off between model accuracy and latency.
In the end, we select the representative configurations of \proj{} and compare it with other sparsity patterns, which demonstrates the acceleration capability of \proj{}.

\subsection{Methodology}

\paragraph{Benchmark.} We evaluate three popular neural networks, VGG (CNN), NMT (LSTM), and BERT (Transformer), which cover tasks from the computer vision and NLP domain.

VGG16~\cite{simonyan2015very} is a popular CNN model with 13 convolutional layers and 3 fully connected layers.
We evaluate its accuracy for image classification on the ImageNet~\cite{krizhevsky2017imagenet} dataset with 1.2 million training images and 50,000 validation images.
\hlc{We prune its weight matrix after applying the im2col method~\mbox{\cite{chetlur2014cudnn}}, which flattens the filters in the same channel to a column and different columns correspond to different channels (so the flattened feature map matrix left multiplies the flattened weight matrix in \mbox{\Fig{fig:tw}}).
This approach is similar to prior work\mbox{~\cite{zhu2019sparse}}.}

We evaluate the accuracy of NMT model, which adopts the attention based encoder-decoder architecture, for the machine translation task~\cite{cho2014properties}.
We reproduce the model with a open-source framework~\cite{luong17}.
We evaluate the NMT model on the IWSLT English-Vietnamese dataset~\cite{luong2016acl_hybrid}, and use the BLEU (bilingual evaluation understudy) score~\cite{papineni2002bleu} as the accuracy metric.

For the state-of-art Transformer model family, we use the BERT-base with $12$ layers.
The two evaluated downstream tasks are the sentence-level classification on the widely used GLUE (general language understanding evaluation) dataset~\cite{wang2019glue} and the more challenging question answering task on the SQuAD dataset~\cite{rajpurkar2016squad,rajpurkar2018know}.
The \hlc{GLUE} dataset is a composite dataset with 10 different sub-tasks, which we evaluate 6 out of them. 

In our experiments, we use the pre-trained models that can achieve their reported reference accuracies.
We then apply \benchmark{EW}, \benchmark{VW}, \benchmark{BW}, and our proposed \benchmark{TW} sparsity patterns to prune the dense models according to the algorithm described in \Sec{sec:pruning_algorithm}.
We use the TensorFlow~\cite{tensorflow2015-whitepaper} framework for fine-tuning. Depending on the dataset size, we perform the fine-tuning for 4-10 epochs at each target sparsity level, which is sufficient to saturate the model accuracy in our experiment.

\paragraph{Baselines.} We compare the proposed \benchmark{TW} with \benchmark{EW}, \benchmark{VW}, and \benchmark{BW}.
For the latency evaluation, we execute \benchmark{EW} and \benchmark{VW} using the cuSparse~\cite{nvidia2019toolkit} library, and execute \benchmark{BW} using the BlockSparse~\cite{blocksparse2020} library released by the authors.
Our \benchmark{TW} implementation (\Sec{sec:implementation}) is based on CUTLASS~\cite{cutlass2019}, an open-source, high-performance GEMM template library. 
\hlc{For all those libraries including \proj{}, we modify the original model codes to explicitly call each library. In the rest of this section, we focus on the GEMM execution time unless explicitly mentioned.}

All the experiments are conducted on the Tesla V100 GPU~\cite{volta2017whitepaper}, which has a peak throughput of 15.7 TFLOPS and 125 TFLOPS for the CUDA cores and tensor cores, respectively. The \benchmark{EW} and \benchmark{VW} run only on the CUDA core with the cuSparse library and the \benchmark{BW} implementation runs only on the tensor core with BlockSparse. 
The convolution operations in the CNN workloads are converted to GEMM by the \benchmark{im2col} method~\cite{chetlur2014cudnn}. The models are all trained using FP32. All inferences on CUDA cores are done using \hlc{FP32}, and all inferences on tensor cores are done using \hlc{FP16}.

\subsection{BERT Results and Design Space Exploration}

We now study the design space of \proj{}, which is the tiling granularity $G$, to explore the trade-off between model accuracy and inference latency.
In addition, we also evaluate the hybrid \benchmark{TEW} pattern, which extends the trade-off space in sparse models.
The analysis is case-studied on the BERT model for sentence pair entailment task on the MNLI dataset.
We report the results of other models/tasks/datasets in the next subsection.

\begin{figure}[t]
    \begin{subfigure}{0.5\columnwidth}
    \includegraphics[width=\linewidth]{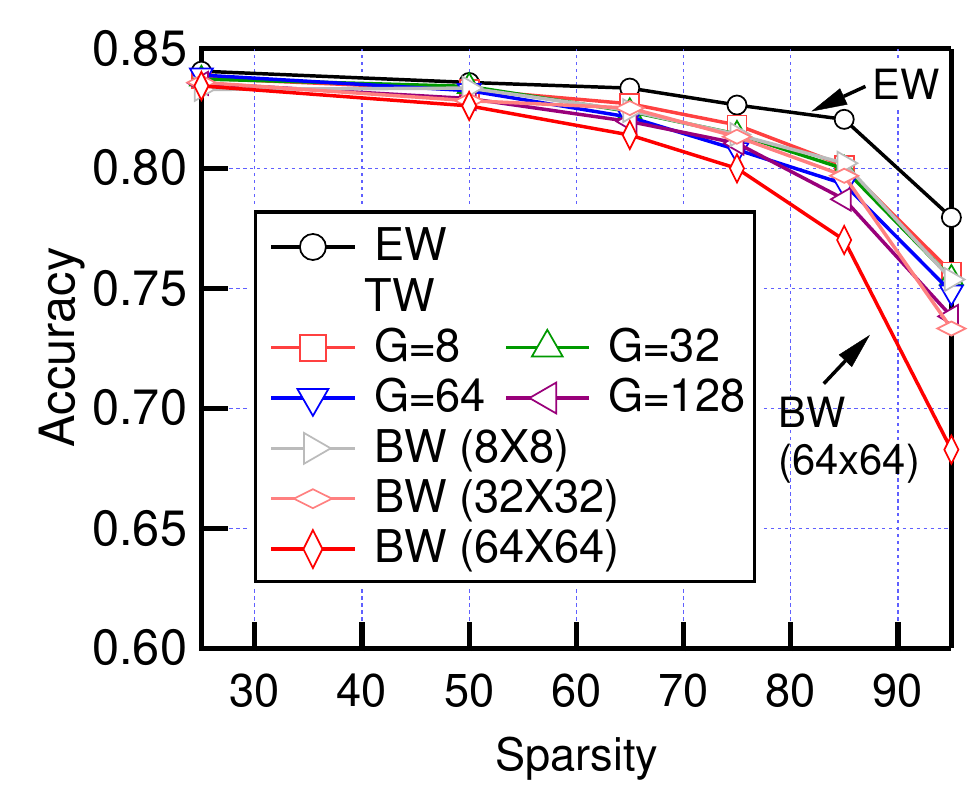}
    \caption{Accuracy.}
    \label{fig:g_accuracy}
    \end{subfigure}~
    \begin{subfigure}{0.5\columnwidth}
    \includegraphics[width=\linewidth]{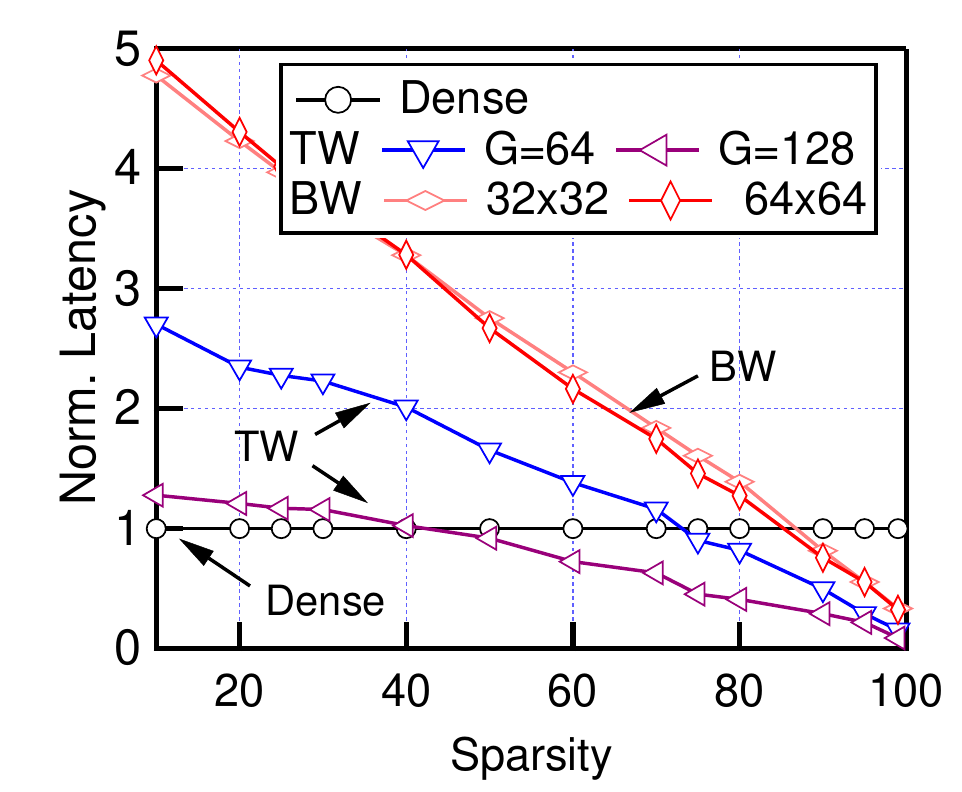}
    \caption{{\hlc{Normalized latency.}}}
        \label{fig:g_latency}
    \end{subfigure}
	\caption{ 
	\hlc{Accuracy and latency of sparse BERT model in \mbox{\benchmark{TW}} and other patterns with different granularities. All latencies are measured on tensor cores and normalized to the dense model.}
	}
	\label{fig:g_impact}
\end{figure}

\paragraph{Impact of \proj{} Granularity.} 
We first explore the impact of tiling granularity $G$ for \benchmark{TW}-based pruning.
\Fig{fig:g_accuracy} compares the accuracy of \benchmark{EW}, \benchmark{BW}, and \benchmark{TW}.
The most fine-grained \benchmark{EW} achieves the best model accuracy as expected.
When sparsity is less than $50\%$, all the granularities evaluated have similar accuracies, suggesting that the BERT model is at least 50\% redundant. In particular, at a sparsity of 75\%, our proposed \texttt{TW} with $G=128$ has an accuracy loss of about 0.9\% and 2.4\% compared to \texttt{EW} and the baseline dense model, respectively. As the sparsity increases, the accuracy drop becomes more significant. The most coarse-grained \benchmark{BW} ($64 \times 64$) experiences the most drastic accuracy drop of 4\% at 75\% sparsity.

The accuracy drop of \texttt{TW} increases slightly with a larger $G$ value. This is because the larger $G$ value puts a more strict constraint on the pruning shape, but larger $G$ also means greater latency reduction.
We find that $G$ of 128 is sufficient to maintain the model accuracy while providing significant latency reduction.
\Fig{fig:g_latency} compares the latency of the dense model, \benchmark{BW}, and \benchmark{TW} on the tensor core.
With only \hlc{40\%} sparsity, \benchmark{TW} with $G=128$ starts to outperform the dense model latency. At a 75\% sparsity, \benchmark{TW-128} achieves a speedup of \hlc{$2.26\times$}.
In contrast, \benchmark{BW} with a block size of $64\times64$ is faster than the dense model only when the sparsity is greater than 90\%, which leads to an accuracy loss as high as 10\% , as shown in \Fig{fig:g_accuracy}.

\begin{figure}[t]
	\hspace*{-.4cm}
    \begin{subfigure}{0.55\columnwidth}
      \includegraphics[width=\linewidth]{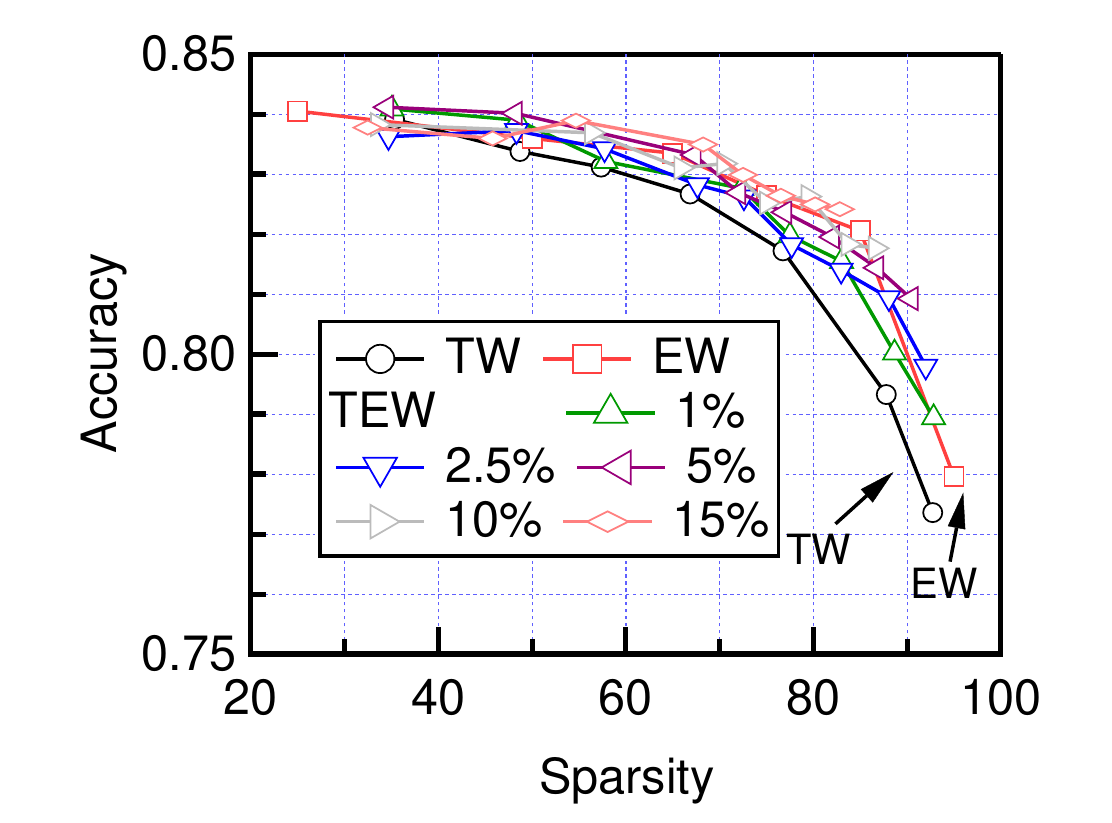}
    \caption{Accuracy.}
    \label{fig:delta_accuracy}
    \end{subfigure}~
	\hspace*{-.4cm}
    \begin{subfigure}{0.55\columnwidth}
             \includegraphics[width=\linewidth]{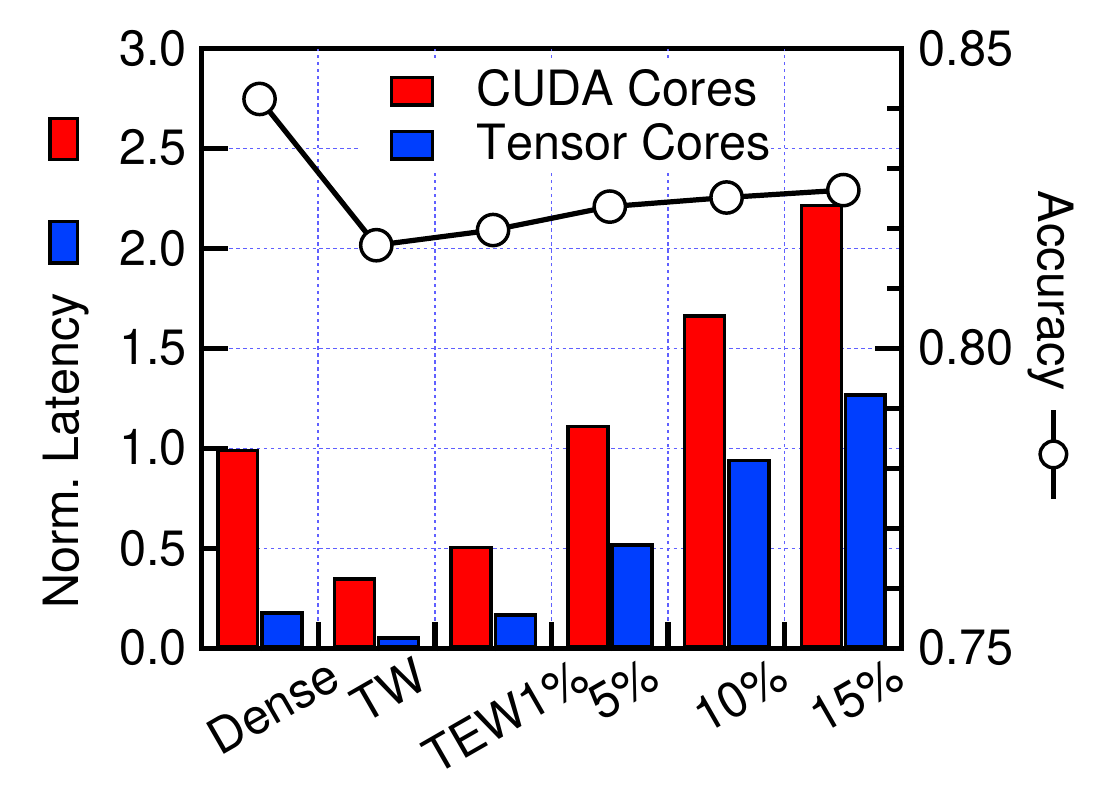}
    \caption{\hlc{Normalized latency.}}
        \label{fig:delta_latency}
    \end{subfigure}
	\caption{Accuracy and latency of \benchmark{TEW}-based sparse BERT model with different $\delta$ values, which determine the portion of added \benchmark{EW} elements. All latency values in (b) are normalized to the latency of dense model on CUDA cores.}
	\label{fig:delta_impact}
\end{figure}

\paragraph{Impact of $\delta$ in \benchmark{TEW}.}
We evaluate the impact of $\delta$ in \benchmark{TEW}, which determines the amount of \benchmark{EW} pattern imposed on \texttt{TW} (\Sec{sec:tile_sparsity}).
\hlc{
\mbox{\Fig{fig:delta_accuracy}} compares the sparse BERT model accuracy of different sparsity levels with \mbox{\benchmark{EW}, \benchmark{TW}, and \benchmark{TEW}} patterns.
The accuracy of the sparse model with \mbox{\benchmark{TW}} is lower than \mbox{\benchmark{EW}}.
}
On the other side, \benchmark{TEW} can mitigate the accuracy loss in \benchmark{TW} by adding a small portion \benchmark{EW} pattern, which is controlled by the $\delta$ parameter in \Sec{subsec:tw}.
For instance, with $\delta = 5\%$, the \benchmark{TEW} accuracy catches up with \benchmark{EW}.

\Fig{fig:delta_latency} compares the latency  (left $y$-axis) and accuracy (right $y$-axis) of the dense model and various \benchmark{TW} and \benchmark{TEW} models with the fixed 75\% sparsity. We show the latency results on both the tensor cores and the CUDA cores, \hlc{which are all normalized to the dense model latency on CUDA cores.}

On the tensor cores, \benchmark{TW} achieves \hlc{$2.26 \times$} speedup than the dense model. \benchmark{TEW} achieves no speedup at $\delta=1\%$ compared to the dense model, and its performance is worse as $\delta$ increases. This is because the irregular portion of \benchmark{TEW} (i.e., the \texttt{EW} portion) could not be executed on the dense tensor cores and, instead, has to be executed on the CUDA Cores, which is about 8$\times$ slower than the tensor cores. To illustrate the point, we show the results of running different sparse models on CUDA cores only.
Using CUDA Cores alone, \benchmark{TEW} with $\delta=1\%$ is about $2\times$ faster than the dense model.
Thus, we expect that \benchmark{TEW} is useful in resource-constraint scenarios such as low-end GPUs with less or even no tensor cores, or mobile systems.

\begin{figure}[t]
	\centering
   % \vspace*{-.3cm}
    \includegraphics[width=1\linewidth]{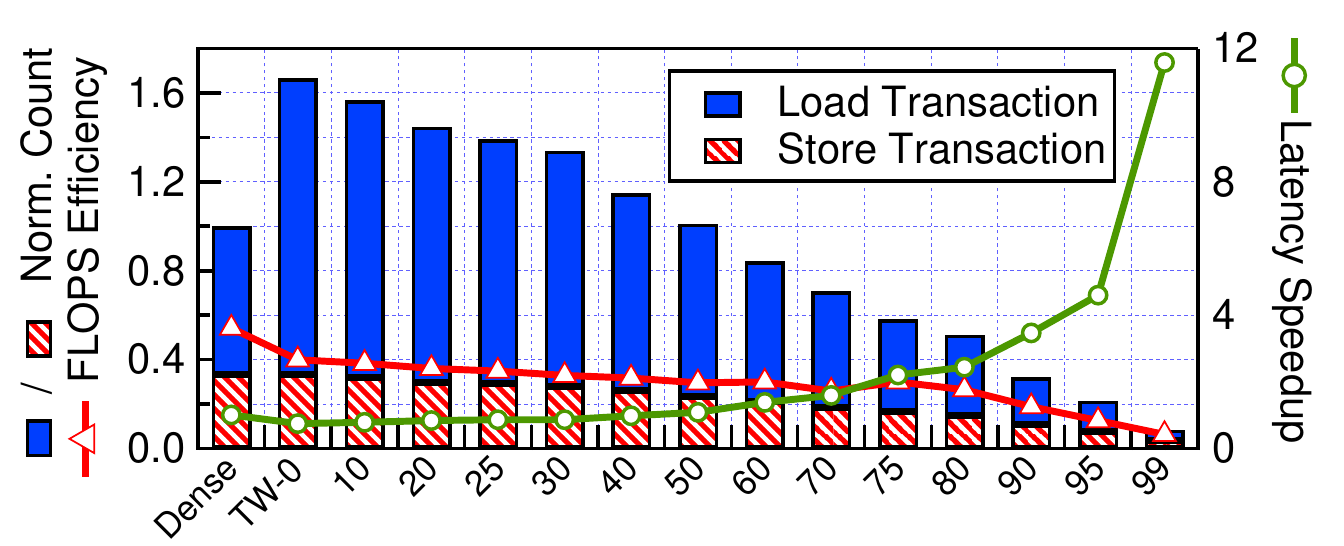}
     \vspace*{-.2cm}
    \caption{The scalability (up to 99\% sparsity) of latency speedup and corresponding performance counters for sparse \benchmark{TW}-based BERT model.
	TW-10 means the model with 10\% \benchmark{TW} sparsity.}
    \label{fig:perf_counter}
    % \vspace*{-.1cm}
    \end{figure}

    \begin{figure*}[t]
      \hspace*{-.1cm}
        \begin{subfigure}{0.5\columnwidth}
        \includegraphics[trim=0 0 0 0, clip, width=\linewidth]{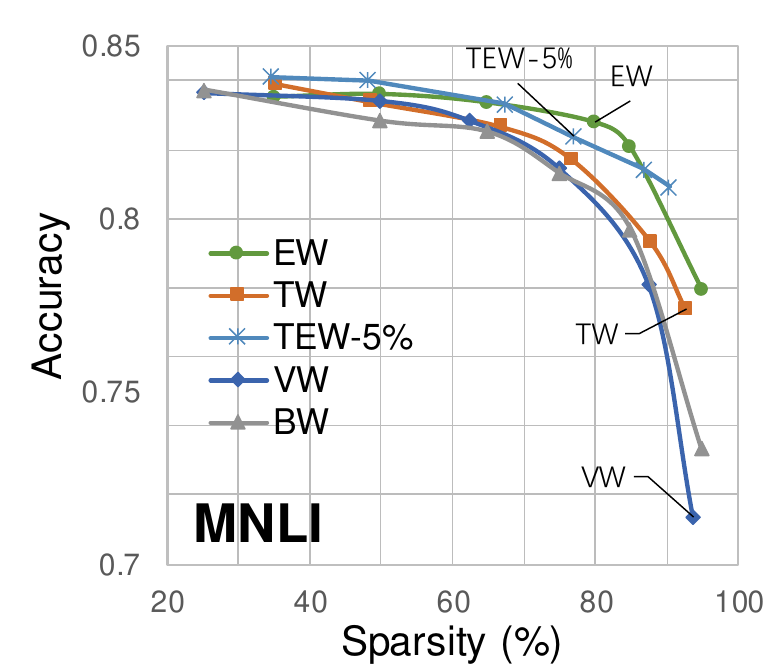}
            %\vspace*{-.2cm}
        \caption{MNLI}
        \label{fig:mnli}
        \end{subfigure}~
        \begin{subfigure}{0.5\columnwidth}
        \includegraphics[trim=0 0 0 0, clip,  width=\linewidth]{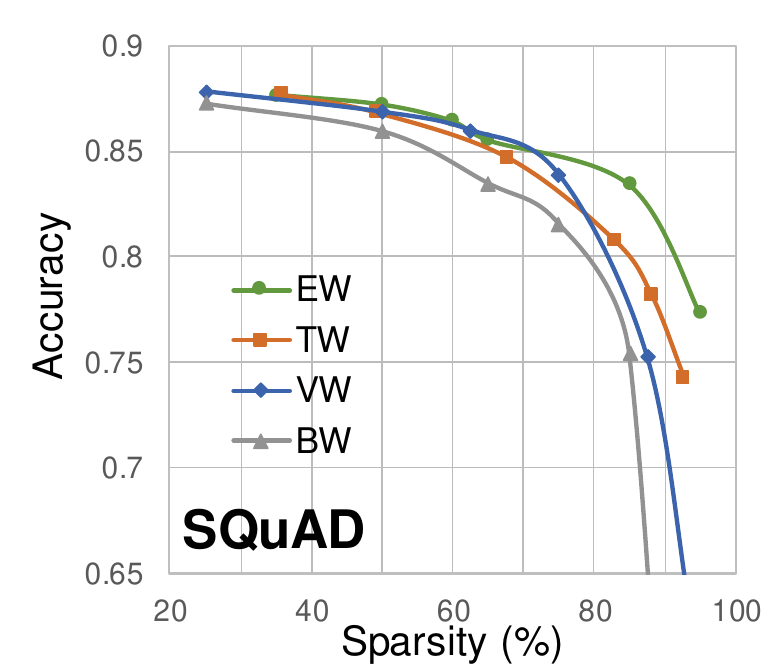}
            %\vspace*{-.2cm}
        \caption{SQuAD}
        \label{fig:squad}
        \end{subfigure}~
        \begin{subfigure}{0.5\columnwidth}
        \includegraphics[trim=0 0 0 0, clip,  width=\linewidth]{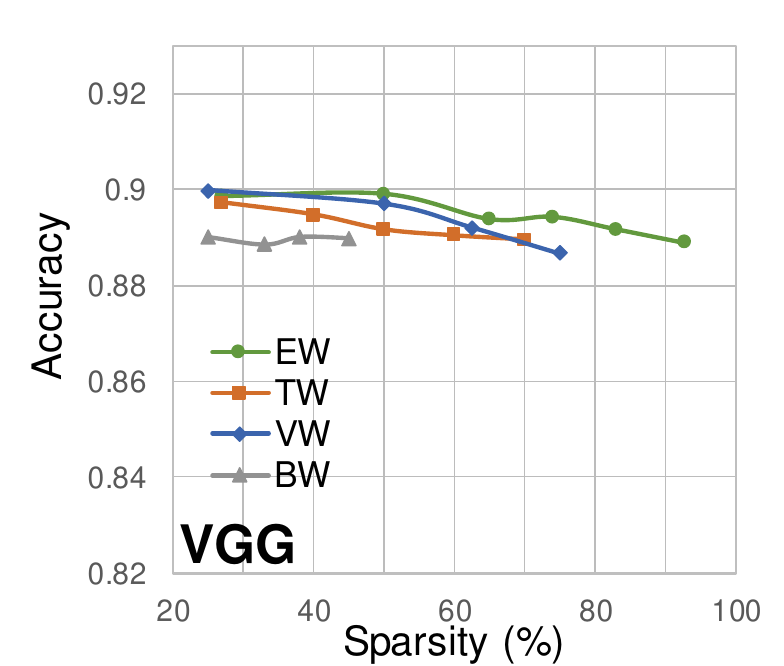}
            %\vspace*{-.2cm}
        \caption{VGG}
        \label{fig:vgg}
        \end{subfigure}~
        \begin{subfigure}{0.47\columnwidth}
        \includegraphics[trim=0 0 0 0, clip, width=\linewidth]{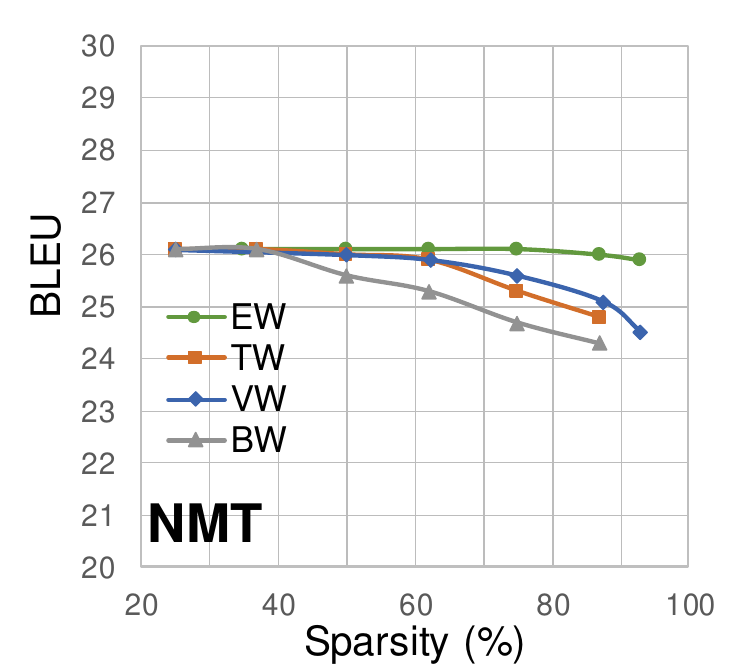}
            %\vspace*{-.2cm}
        \caption{NMT}
        \label{fig:nmt}
        \end{subfigure}
        \caption{The accuracy comparison of different models on the model-specific downstream tasks with various pruning patterns and varying sparsity levels. Plot (a) and (b) are BERT models evaluated on GLUE dataset and SQuAD dataset, respectively. Plot (c) is VGG16 model evaluated on ImageNet. Plot (d) is LSTM model evaluated on NMT task.}
        \label{fig:accuracy_all}
       % \vspace*{-.2cm}
    \end{figure*}
    
\paragraph{Speedup Scalability.}
	We also study the speedup scalability by intentionally pruning DNN models to an extreme sparsity level.
	It is highly likely that the size of DNN models would continue to \mbox{grow~\cite{brown2020language}}. 
	\mbox{\Fig{fig:perf_counter}} shows the latency speedup of sparse BERT model with \mbox{\benchmark{TW}} on tensor cores until 99\% sparsity.
	At the 99\% sparsity level, \mbox{\benchmark{TW}} with $G=128$ achieves $11.6\times$ speedup, demonstrating its significant acceleration potential.

\paragraph{\hlc{Performance Counters.}}
\hlc{
\mbox{\Fig{fig:perf_counter}} also shows the total number of global memory load/store requests and FLOPS (floating operations per second) efficiency for running the sparse \mbox{\benchmark{TW}}-based BERT model, which are all normalized to the dense BERT model.
Compared to the original BERT model, our \mbox{\benchmark{TW}} implementation with zero sparsity generates twice of global memory request owing to the masking overhead, for which we use the int32 format.
The extra load traffic leads to about 35\% performance loss. 
With about 40\% sparsity, the benefit outperforms the overhead, leading to the net latency speedup.
The FLOPS efficiency equals the measured FLOPS divided by all tensors' peak FLOPS. 
The sparse \mbox{\benchmark{TW}} model maintains a relatively high FLOPS efficiency until 80\% sparsity and quickly drops after that owing to the reduced computation demand. 
}

\subsection{Comparison with Other Patterns}

We compare the accuracy and latency speedup of \benchmark{TW} with \benchmark{EW}, \benchmark{VW}, \benchmark{BW} on three different models. 
We perform the comprehensive evaluation of BERT model for the sentence classification task on the composite GLUE dataset, which includes ten different datasets.
We observe similar results on 6 studied datasets and therefore only report the result on the largest dataset MNLI.
We also report its result on the question answering task with the SQuAD dataset.
For the latency speedup, we report the results on the V100 GPU using tensor cores and CUDA cores separately.

\paragraph{Accuracy.}
\Fig{fig:accuracy_all} shows the accuracy of different models with different pruning patterns.
The granularity of \proj{} is 128 and block size for \benchmark{BW} is $32 \times 32$, which balances the accuracy and latency speedup as our previous design space analysis suggests.
The vector size of \benchmark{VW} is set to 16 as used in the original paper~\cite{zhu2019sparse}.
\benchmark{EW} reaches the best accuracy of all the evaluated algorithms and \benchmark{BW} has the worst accuracy under the same sparsity.
The accuracy of \benchmark{TW} and \benchmark{VW} are similar when the sparsity is below 70\%. 
With high sparsity ($> 70\%$), \benchmark{TW} generally outperforms the \benchmark{VW} with the exception of NMT. 

\benchmark{TW} achieves better accuracy when sparsity is high because it allows the uneven distribution sparsity in a weight matrix.
\Fig{fig:pattern} shows the resulted weight sparsity distributions of layer 0 of BERT under the $75\%$ sparsity for different patterns.
The \benchmark{EW} result shows that there exits uneven distribution across the matrix.
And \benchmark{VW} is unable to fit this characteristic because it forces all prune units (vector) to have the same sparsity.
In contrast, both \benchmark{BW} and \benchmark{TW} can adapt to this sparsity locality.
Meanwhile,\benchmark{VW} cannot adapt to the uneven sparsity distribution across different layers as explained in \Sec{sparsity:comparison} while \benchmark{TW} can.
For the NMT model, both \benchmark{VW} and \benchmark{TW} experience a rapid accuracy drop compared to \benchmark{EW} when the sparsity is over 60\%, which suggests this model prefers irregular sparsities.
\benchmark{VW} slightly outperforms \benchmark{TW} owing to its smaller granularity of 16.

\begin{figure}[t]
 %\centering
 \vspace*{-0.3cm}
 \hspace*{0.9cm}
%  \hspace{-0.3cm}
    \begin{subfigure}{0.33\columnwidth}
    \includegraphics[width=\linewidth]{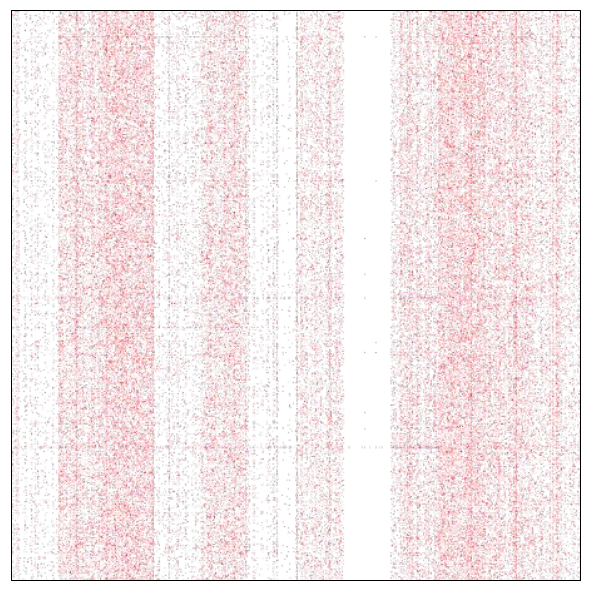} 
    % \vspace*{-1cm}
    \caption{\benchmark{EW}}
    \vspace*{0.3cm}
    \label{fig:pattern_a}
    \end{subfigure}~
    \hspace{0.5cm}
    \begin{subfigure}{0.33\columnwidth}
    \includegraphics[width=\linewidth]{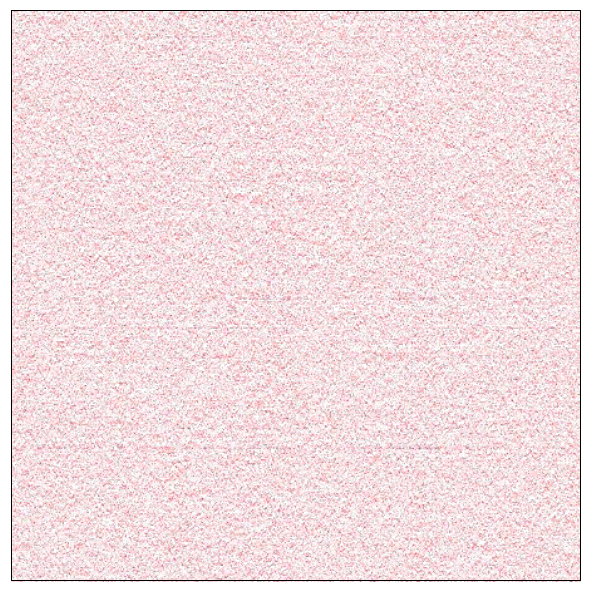}
    % \vspace*{-1cm}
    \caption{\benchmark{VW}}
    \vspace*{0.3cm}
    \label{fig:pattern_b}
    \end{subfigure}~
    \vfill
    \hspace*{0.9cm}
    \begin{subfigure}{0.33\columnwidth}
    \includegraphics[width=\linewidth]{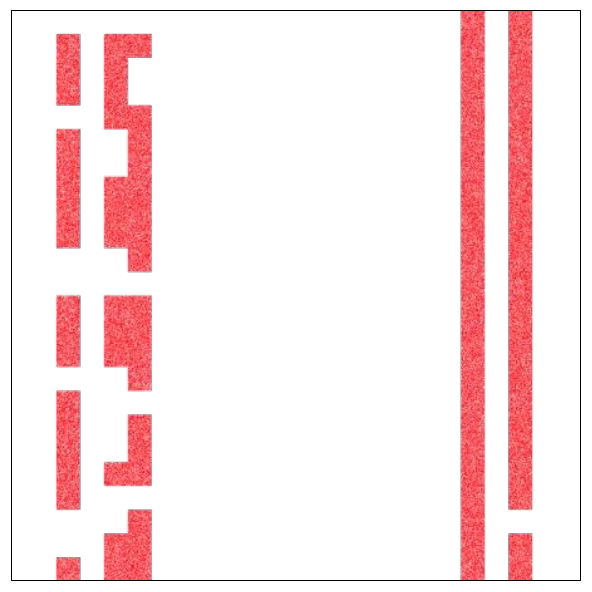} 
    % \vspace*{-1cm}
    \caption{\benchmark{BW}}
    \label{fig:pattern_c}
    \end{subfigure}~
    \hspace{0.5cm}
    \begin{subfigure}{0.33\columnwidth}
    \includegraphics[width=\linewidth]{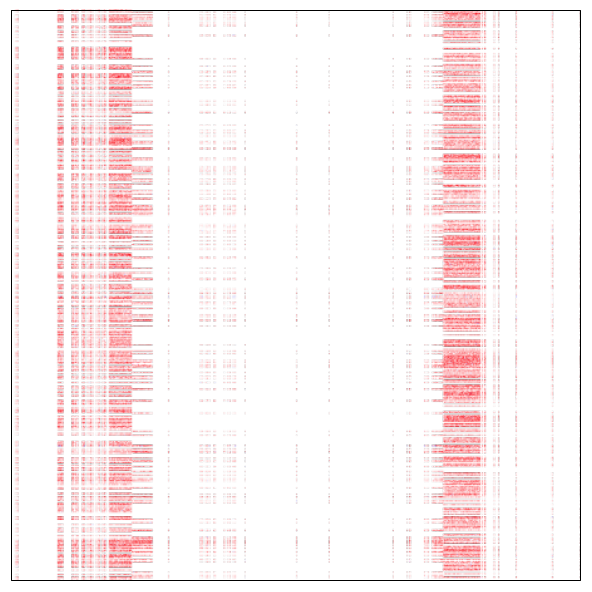}
    % \vspace*{-1cm}
    \caption{\benchmark{TW} }
    \label{fig:pattern_d}
    \end{subfigure}     
    %  \vspace*{-0.3cm}
    \caption{Different pruning patterns under 75\% sparsity on layer 0 attention matrix $\omega_Q$ in BERT model.}
    \label{fig:pattern}
\end{figure}

\begin{figure}[t]
 \vspace*{-0.4cm}
    \hspace*{0.3cm}
    \begin{subfigure}{\linewidth}
	\includegraphics[trim=10 0 30 0, clip, width=.45\linewidth]{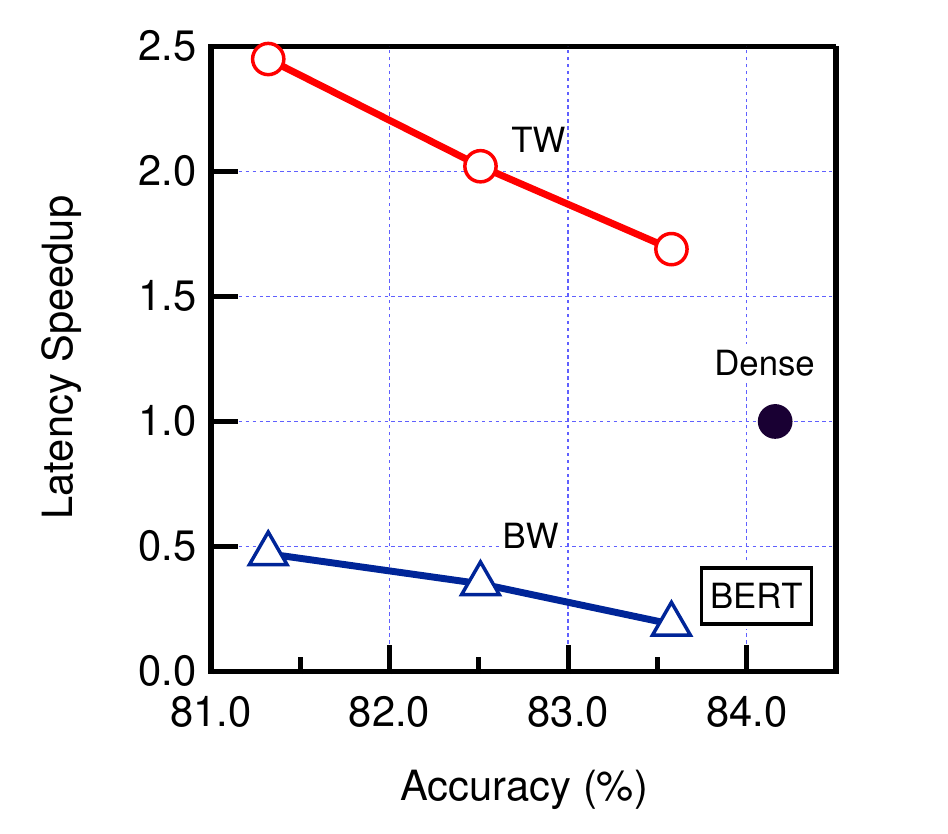}
    \includegraphics[trim=10 0 30 0, clip, width=.45\linewidth]{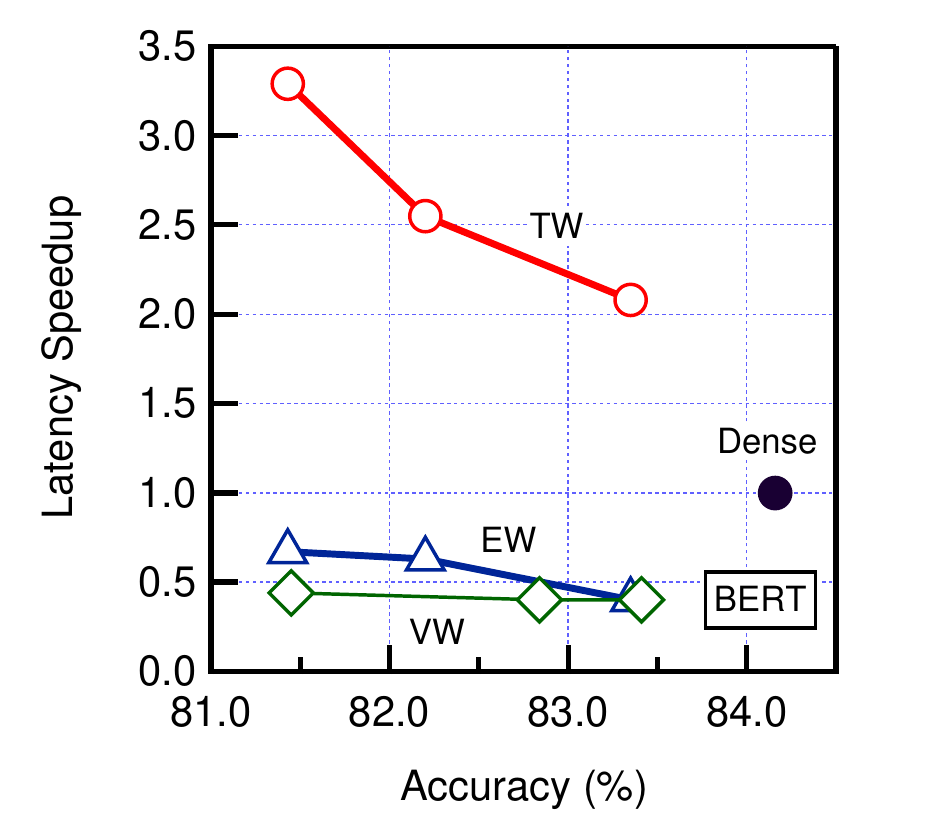}
    \vspace*{-0.3cm}
    \caption{BERT}
    \vspace*{0.1cm}
	\label{fig:speedup_accuracy_bert}
	\end{subfigure}
    
    \hspace*{0.3cm}
    \begin{subfigure}{\linewidth}
	\includegraphics[trim=10 0 30 0, clip, width=.45\linewidth]{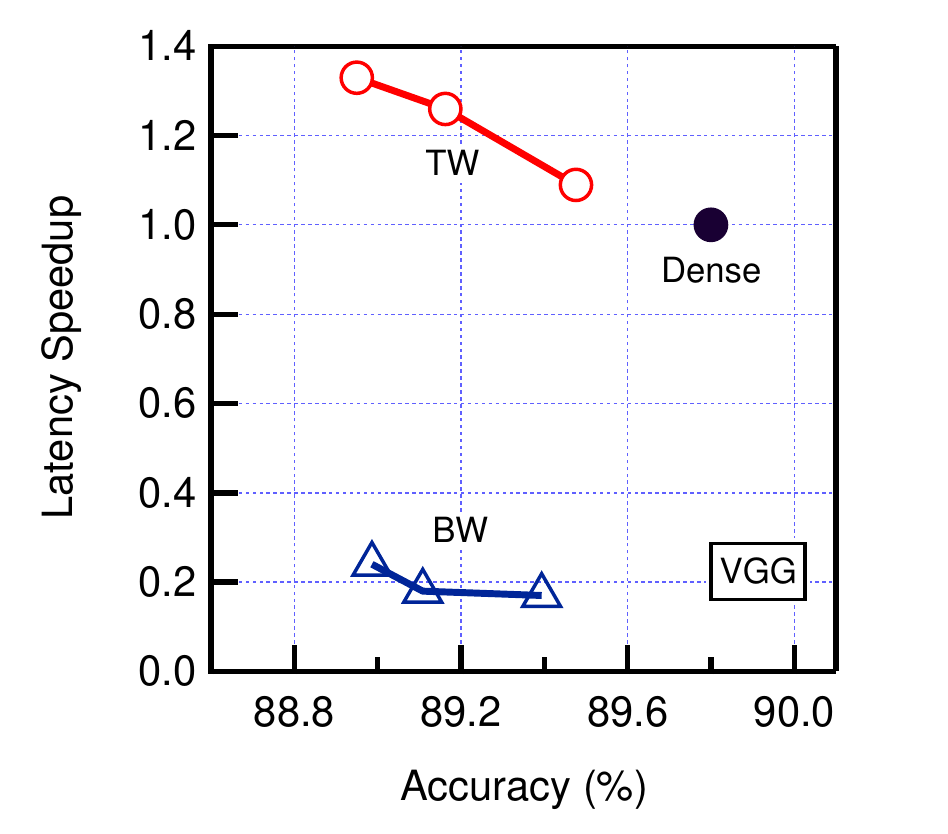}
    \includegraphics[trim=10 0 30 0, clip, width=.45\linewidth]{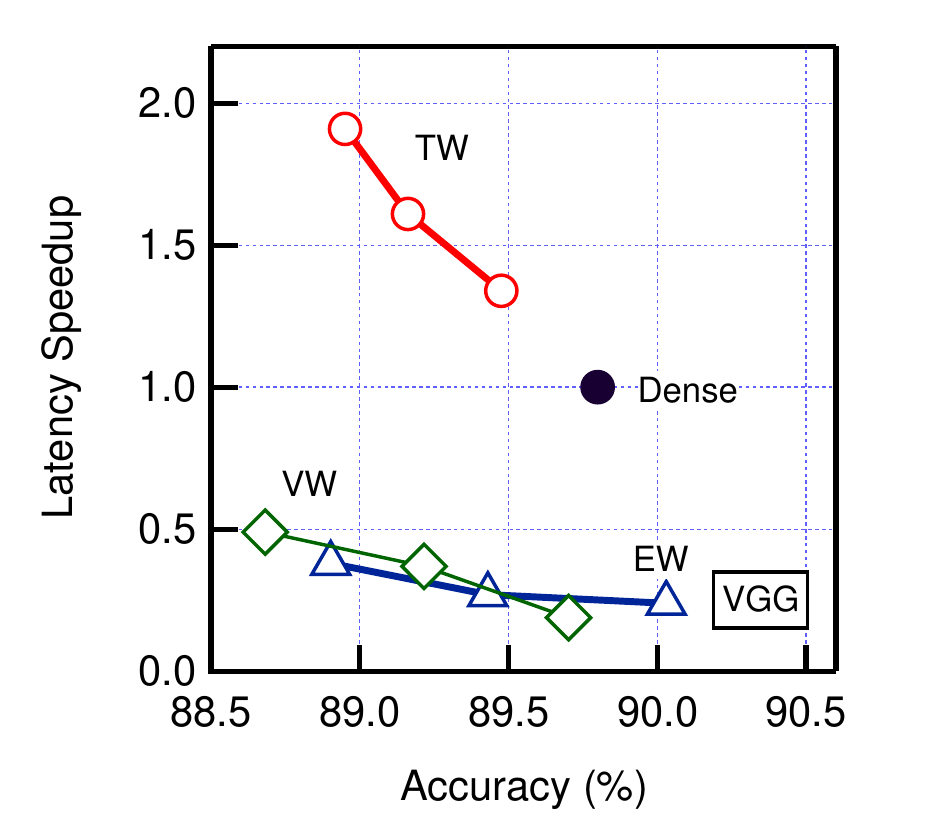}
    \vspace*{-0.3cm}
    \caption{VGG}
    \vspace*{0.1cm}
	\label{fig:speedup_accuracy_vgg}
	\end{subfigure}
    
    \hspace*{0.3cm}
    \begin{subfigure}{\linewidth}
	\includegraphics[trim=10 0 30 0, clip, width=.45\linewidth]{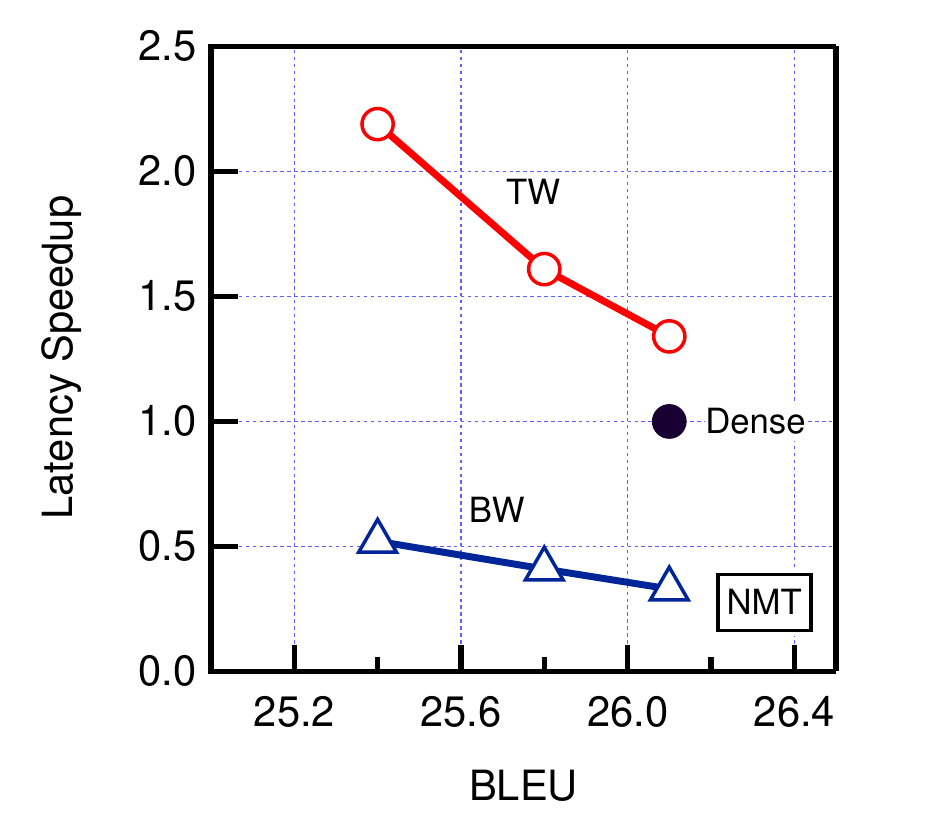}
    \includegraphics[trim=10 0 30 0, clip, width=.45\linewidth]{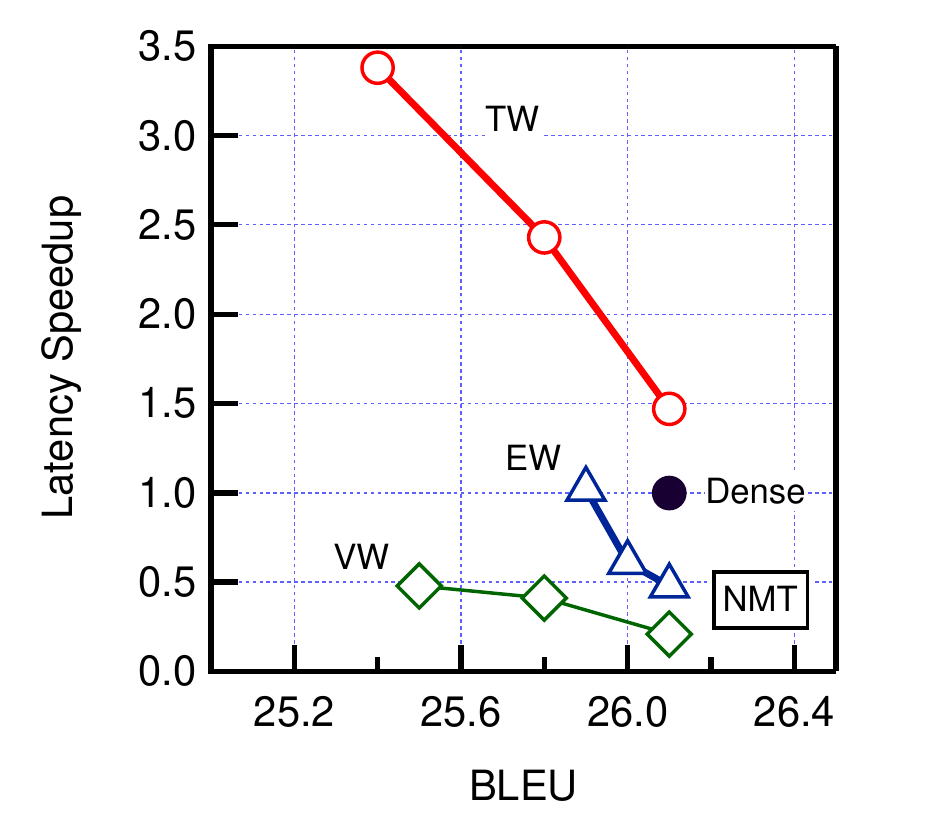}
    \vspace*{-0.3cm}
    \caption{NMT}
    \vspace*{0.1cm}
	\label{fig:speedup_accuracy_NMT}
	\end{subfigure}
	 \caption{The trade-off between latency speedup and model accuracy. The speedup is calculated on tensor cores (left column) and CUDA cores (right column) separately.}
	\label{fig:speedup_accuracy}
\end{figure}

\paragraph{Speedup vs Accuracy.}
\Fig{fig:speedup_accuracy} compares the trade-off of latency speedup and model accuracy based on \benchmark{TW} and other patterns including \benchmark{BW}, \benchmark{VW}, and \benchmark{EW}. In specific, we compare the \benchmark{TW} and \benchmark{BW} running on the tensor cores, and compare the \benchmark{TW}, \benchmark{VW}, and \benchmark{EW} on the CUDA cores. The speedup is calculated against dense models on the tensor cores and CUDA cores separately. 
    The experimental results demonstrate that only \benchmark{TW} can extend the latency-accuracy Pareto frontier on tensor cores and CUDA cores. In contrast, other sparsity patterns lead to both longer latency and lower accuracy than the dense model.

    Finally, we compare the latency speedup of various patterns with the same level of accuracy drop (BERT with $< 3\%$ drop, VGG with $< 1\%$ drop and NMT with $< 1$ BLEU drop).
    On tensor cores, \proj{} achieves an average speedup of $1.95\times$ while \benchmark{BW} is $0.41\times$.
    On CUDA cores, \proj{} achieves an average speedup of $2.86 \times$ while \benchmark{EW} and \benchmark{VW} are $0.69 \times$ and $0.47\times$.
    \proj{} achieves the meaningful latency reduction on both tensor cores and CUDA cores owing to its compatibility with dense GEMM, while all other sparsity patterns cause the actual slowdown.

\subsection{End-to-end Latency and Impact of Optimizations}

The above performance comparison between the dense model and various sparse models only considers the GEMM-related computation. 
We now study the end-to-end latency and impact of optimizations presented in \Sec{sec:implementation}, which includes the transposed matrix storage and kernel fusion.
\Fig{fig:end_to_end_breakdown} shows the end-to-end latency breakdown with different optimization combinations when running the sparse \benchmark{TW} BERT model with 75\% sparsity. 
We do not use the VGG in this experiment because it only includes 5\% non-GEMM computations.
Without performing the matrix transpose optimization, the GEMM computation cannot benefit from the high sparsity.
The transpose kernel takes about 10\% of overall latency without fusion.
With the transpose and fusion, the GEMM-only speedup for BERT and NMT is $2.26\times$ and $2.38\times$ respectively, while the end-to-end speedup is $1.61\times$ and $1.86\times$ on the tensor core.

    \begin{figure}[t]
    \centering
    \includegraphics[width=.9\columnwidth]{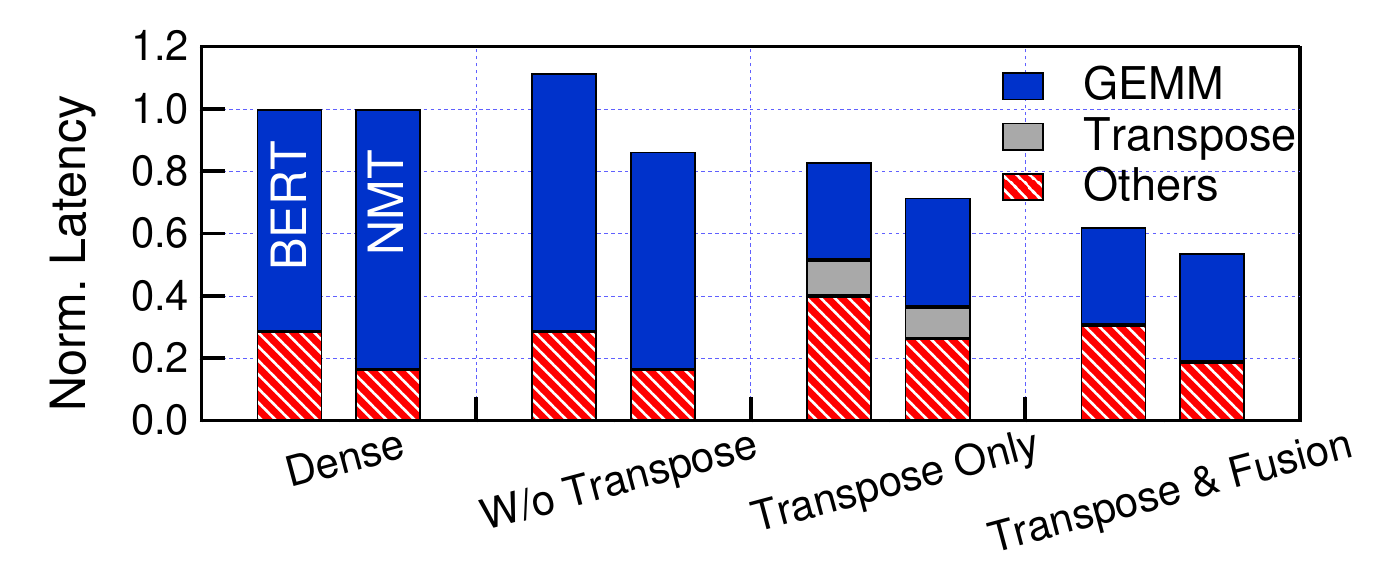}
    \caption{The end-to-end latency breakdown for sparse \benchmark{TW}-based BERT and NMT with 75\% sparsity ($<3\%$ accuracy drop and $<1$ BLEU drop than the dense model respectively). We enable the transpose and fusion optimization by default.}
    \label{fig:end_to_end_breakdown}
\end{figure}

\section{Related Work and Discussion}
\label{sec:related_work}

\paragraph{Architectural Support for Sparse DNN.}
Recently, designing efficient architectures for sparse DNN models has become an active research topic. There are many prior works to dealing with sparsity through architectural support~\cite{hua2019boosting, zhang2019eager, deng2019tie, yang2019sparse, hegde2019extensor, sadi2019efficient}. ExTensor~\cite{hegde2019extensor} proposes a novel approach to accelerate tensor algebra kernels using the principle of hierarchical computation elimination in the presence of sparsity. Sparse ReRAM Engine~\cite{yang2019sparse} exploits both the weight and activation sparsity to accelerate the DNN model. The channel gating~\cite{hua2019boosting} is a fine-grained dynamic pruning technique for CNN inference. Those work are all based on customized ASIC or FPGA. 
There are also prior works~\cite{guo2020balancing_arxiv,zhu2019sparse} that focus on optimizing the tensor core for better performance or flexibility.

\paragraph{Software Optimization for Sparse DNN.}
There are also previous works that propose software optimization for sparse model acceleration on modern architectures without hardware modification.
Prior work proposes an efficient mixed-mode representation called MM-CSF for sparse tensor, which partitions nonzero elements into disjoint sections for performance acceleration~\cite{nisa2019efficient}. 
CFS SpMV is a new optimization strategy for the sparse matrix-vector multiplication and shows high performance on multicore architectures~\cite{elafrou2019conflict}.  DCSR (densified compressed sparse row) is well-suited to GPU architecture for sparse computation using the near-memory strategy~\cite{fujiki2019near}.

\paragraph{\proj{} on Other Platforms.}
 Although we only implement \proj{} on the GPU platform, it is quite possible to support our sparsity pattern on other platforms like TPU~\cite{jouppi2017datacenter}. 
    The fundamental requirement of supporting \proj{} is the medium size GEMM.
Our evaluation shows that \proj{} with $G=128$ strikes a balance between model accuracy and latency, which implies the requirement of $128\times N \times128$ GEMM.
    The latest TPU~\cite{tpuv2} adopts a relatively large systolic array ($128 \times 128$), which meets the aforementioned requirement.
    However, it only exposes high-level programming interfaces like GEMM, which makes the other optimization like streaming concurrency difficult. 
    In other words, supporting \proj{} on other platforms like TPU is feasible if their low-level programming interfaces are exposed.
    
\paragraph{Sparsity Patterns.}
The sparsity pattern plays an important role in both the model accuracy and architecture design for sparse DNNs.
Zhu et al. propose the vector-wise pruning pattern and the corresponding sparse tensor core architecture~\cite{zhu2019sparse}.
The vector-wise pruning pattern adds constraints on the sparsity of each pruning unit to guarantee the pruned matrices to be acceleration-friendly.
They reported accuracy results on popular CNN and RNN models.
However, this method fails to capture the uneven sparsity distribution across different model layers, which limits its pruning effect.
Narang et al. propose the block-wise pruning pattern~\cite{narang2017block}.
This pattern has the prune unit as a block, making it execution-friendly on the dense GPU architecture.
However, their method has a strong constraint on the pruning shape which impacts the model accuracy significantly.

Recent work also explores energy-oriented pruning, targeting both accelerators~\cite{yang2018energy, yang2017designing} and general-purpose processors~\cite{yang2018ecc, yang2018netadapt}. Our work removes redundant computations and thus could also reduce energy consumption. Yang et al.~\cite{yang2020automatic} demonstrate quantization-pruning joint compression; we leave it to future work to explore how to integrate \textit{tile sparsity} with quantization.

\section{Conclusion}\label{sec:conclude}

In this work, we propose to co-design the tiling of matrix multiplication and DNN model pruning pattern, with the purpose of balancing the irregularity for the model accuracy and compatibility for dense GEMM computation.
We study an efficient software-only implement of our proposed sparsity pattern, \proj, that leverages the tensor core accelerator and concurrency features in the GPU.
We demonstrate its capability of model accuracy preserving and high performance speedup on the state-of-the-art DNN models.

\section*{Acknowledgement}

We thank the anonymous reviewers for their constructive feedback for improving the work.
This work was supported by National Key R\&D Program of China (2019YFF0302600), the National Natural Science Foundation of China (NSFC) grant (61702328 and 61832006). Any opinions, findings, and conclusions in this paper are those of the authors only and do not necessarily reflect the views of our sponsors. 
%-------------------------------------------------------------------------------
\bibliographystyle{IEEEtranS}
\bibliography{references}

%%%%%%%%%%%%%%%%%%%%%%%%%%%%%%%%%%%%%%%%%%%%%%%%%%%%%%%%%%%%%%%%%%%%%%%%%%%%%%%%
\end{document}